\documentclass[journal,12pt,onecolumn]{IEEEtran}
\IEEEoverridecommandlockouts
\ifCLASSINFOpdf
\else
\fi

\usepackage{snapshot}
\usepackage{graphicx}
\usepackage{latexsym,bm}
\usepackage{url}
\pdfoutput=1
\usepackage{amsmath}
\usepackage{amssymb}
\usepackage{amsthm}
\usepackage{epstopdf}
\usepackage{cite}
\usepackage{multirow}
\usepackage{subcaption}
\usepackage[mathscr]{euscript}

\usepackage{algorithm}
\usepackage{algpseudocode}

\makeatletter
\def\therule{\makebox[\algorithmicindent][l]{\hspace*{.5em}\vrule height .75\baselineskip depth .25\baselineskip}}%

\newtoks\therules
\therules={}
\def\appendto#1#2{\expandafter#1\expandafter{\the#1#2}}
\def\gobblefirst#1{
  #1\expandafter\expandafter\expandafter{\expandafter\@gobble\the#1}}%
\def\LState{\State\unskip\the\therules}
\def\pushindent{\appendto\therules\therule}%
\def\popindent{\gobblefirst\therules}%
\def\printindent{\unskip\the\therules}%
\def\printandpush{\printindent\pushindent}%
\def\popandprint{\popindent\printindent}%

\algdef{SE}[WHILE]{While}{EndWhile}[1]
  {\printandpush\algorithmicwhile\ #1\ \algorithmicdo}
  {\popandprint\algorithmicend\ \algorithmicwhile}%
\algdef{SE}[FOR]{For}{EndFor}[1]
  {\printandpush\algorithmicfor\ #1\ \algorithmicdo}
  {\popandprint\algorithmicend\ \algorithmicfor}%
\algdef{S}[FOR]{ForAll}[1]
  {\printindent\algorithmicforall\ #1\ \algorithmicdo}%
\algdef{SE}[LOOP]{Loop}{EndLoop}
  {\printandpush\algorithmicloop}
  {\popandprint\algorithmicend\ \algorithmicloop}%
\algdef{SE}[REPEAT]{Repeat}{Until}
  {\printandpush\algorithmicrepeat}[1]
  {\popandprint\algorithmicuntil\ #1}%
\algdef{SE}[IF]{If}{EndIf}[1]
  {\printandpush\algorithmicif\ #1\ \algorithmicthen}
  {\popandprint\algorithmicend\ \algorithmicif}%
\algdef{C}[IF]{IF}{ElsIf}[1]
  {\popandprint\pushindent\algorithmicelse\ \algorithmicif\ #1\ \algorithmicthen}%
\algdef{Ce}[ELSE]{IF}{Else}{EndIf}
  {\popandprint\pushindent\algorithmicelse}%
\algdef{SE}[PROCEDURE]{Procedure}{EndProcedure}[2]
   {\printandpush\algorithmicprocedure\ \textproc{#1}\ifthenelse{\equal{#2}{}}{}{(#2)}}%
   {\popandprint\algorithmicend\ \algorithmicprocedure}%
\algdef{SE}[FUNCTION]{Function}{EndFunction}[2]
   {\printandpush\algorithmicfunction\ \textproc{#1}\ifthenelse{\equal{#2}{}}{}{(#2)}}%
   {\popandprint\algorithmicend\ \algorithmicfunction}%
\makeatother

\begin{document}
\begin{titlepage}
\begin{center}
\vspace*{-2\baselineskip}
\begin{minipage}[l]{7cm}
\flushleft
\includegraphics[width=2 in]{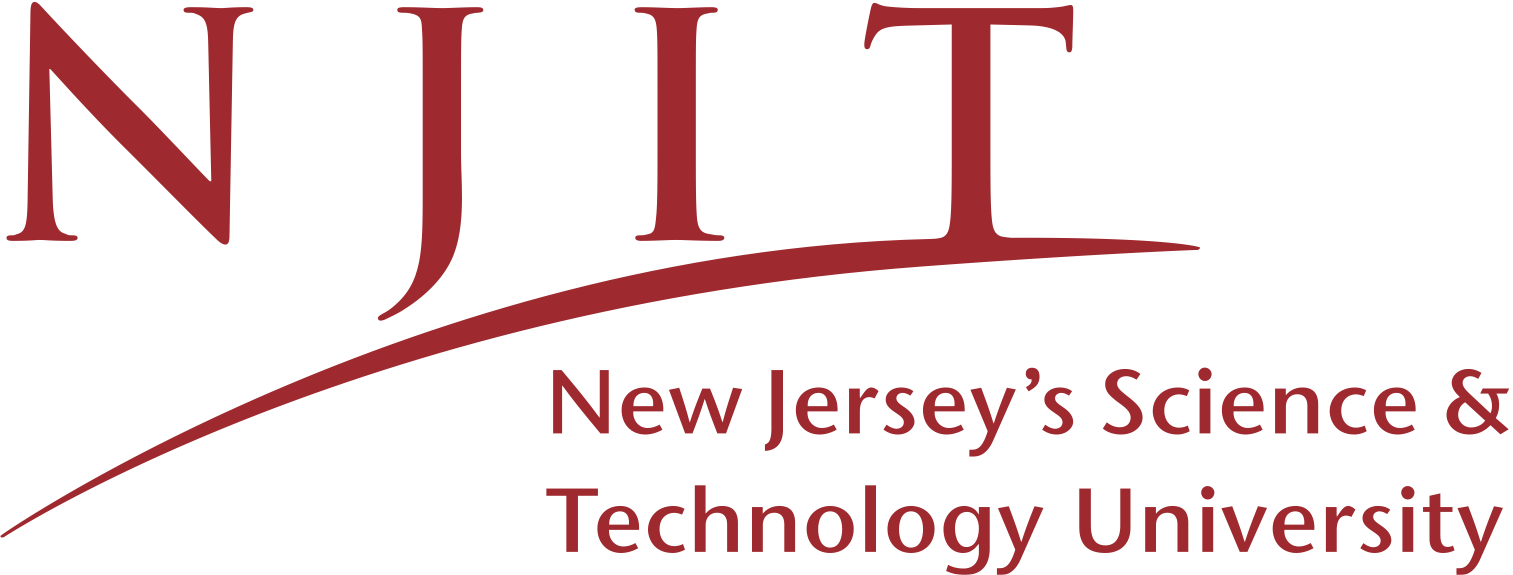}
\end{minipage}
\hfill
\begin{minipage}[r]{7cm}
\flushright
\includegraphics[width=1 in]{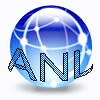}%
\end{minipage}

\vfill

\textsc{\LARGE Joint Spectrum and Power Allocation for  \\[12pt]
Multi-node Cooperative Wireless Systems}\\

\vfill
\textsc{\LARGE XUEQING HUANG\\[12pt]
\LARGE NIRWAN ANSARI}\\
\vfill
\textsc{\LARGE TR-ANL-2014-002\\[12pt]
\LARGE May 22, 2014}\\[1.5cm]
\vfill
{ADVANCED NETWORKING LABORATORY\\
 DEPARTMENT OF ELECTRICAL AND COMPUTER ENGINEERING\\
 NEW JERSY INSTITUTE OF TECHNOLOGY}
\end{center}
\end{titlepage}

\begin{abstract}
Energy efficiency is a growing concern for wireless networks, not only due to the emerging traffic demand from smart devices, but also because of the dependence on the traditional unsustainable energy and the overall environmental concerns. The urgent call for reducing power consumption while meeting system requirements has motivated increasing research efforts on green radio. In this paper, we investigate a new joint spectrum and power allocation scheme for a cooperative downlink multi-user system using the frequency division multiple access scheme, in which arbitrary $M$ base stations (BSs) coordinately allocate their resources to each user equipment (UE). With the assumption that multi-BS UE (user being served by multi-BS) would require the same amount of spectrum from these BSs, we conclude that when the number of multi-BS UEs is limited by $M-1$, the resource allocation scheme can always guarantee the minimum overall transmit power consumption while meeting the throughput requirement of each UE and also each BS's power constraint. Then, to decide the clusters of multi-BS UEs and the clusters of individual-BS UEs (users being served by individual BSs), we propose a UE-BS association scheme and a complexity reduction scheme. Finally, a novel joint spectrum and power allocation algorithm is proposed to minimize the total power consumption. Simulation results are presented to verify the optimality of the derived schemes.
\end{abstract}

\begin{keywords}
Joint spectrum and power allocation, cooperative transmission, frequency division multiple access, UE-BS association, complexity reduction.
\end{keywords}

\section{Introduction}
To improve the system performance and make the best use of the system resource, cooperative transmissions have recently attracted much attention\cite{Capacity}, \cite{CooMp}. The basic idea is to take advantage of the broadcast nature of wireless communications such that some nodes in wireless networks can help each other to transmit signals for better quality via spatial diversity or higher data rates through spatial multiplexing. 

\begin{figure*}
\centering
\hspace*{\fill}
\begin{subfigure}[b]{2 in}
\includegraphics[width=\textwidth]{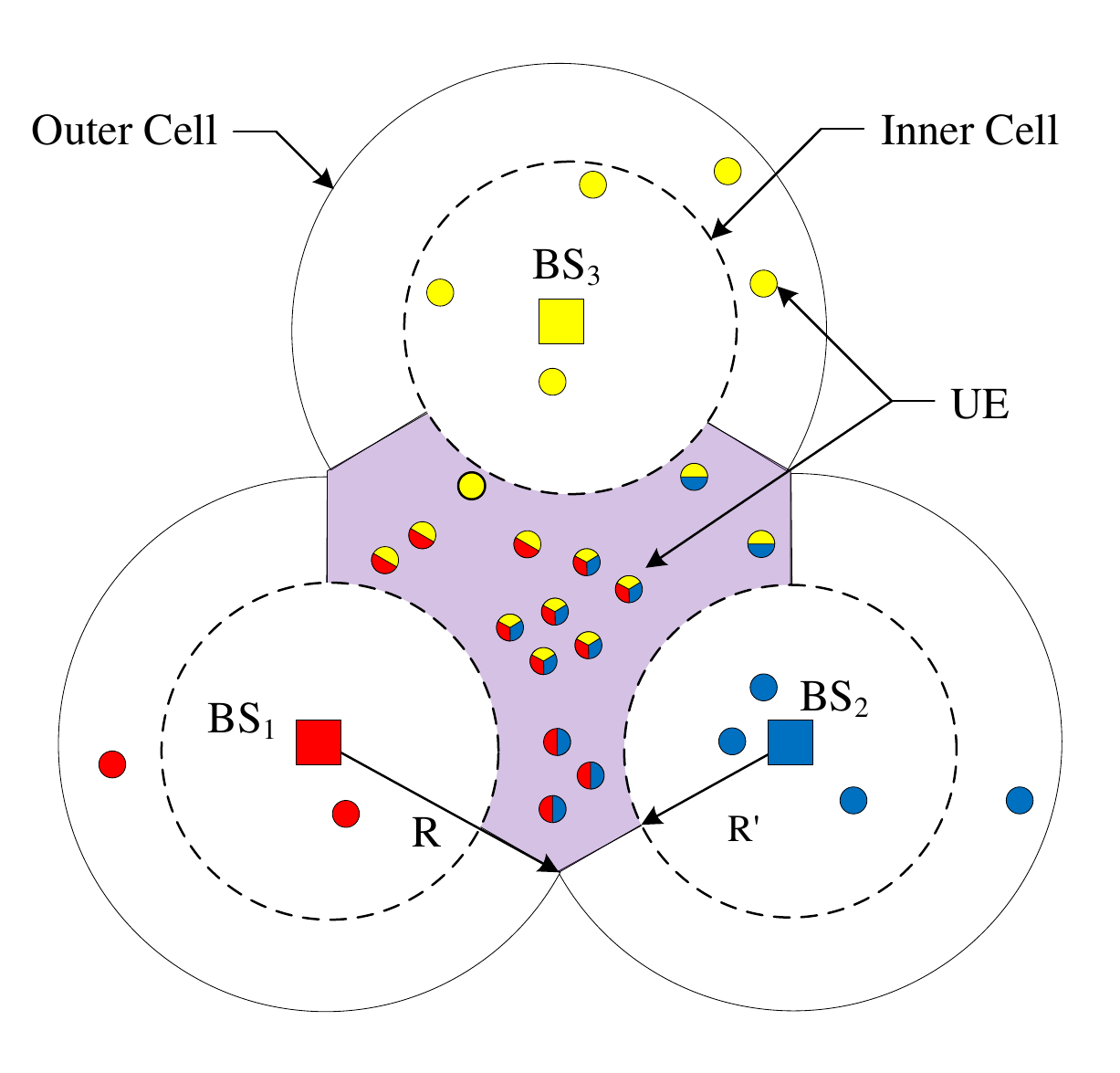}
\caption{Color blocks in UE indicate serving BS candidates}
\label{cellular}
\end{subfigure}\hfill
\begin{subfigure}[b]{2.5 in}
\includegraphics[width=\textwidth]{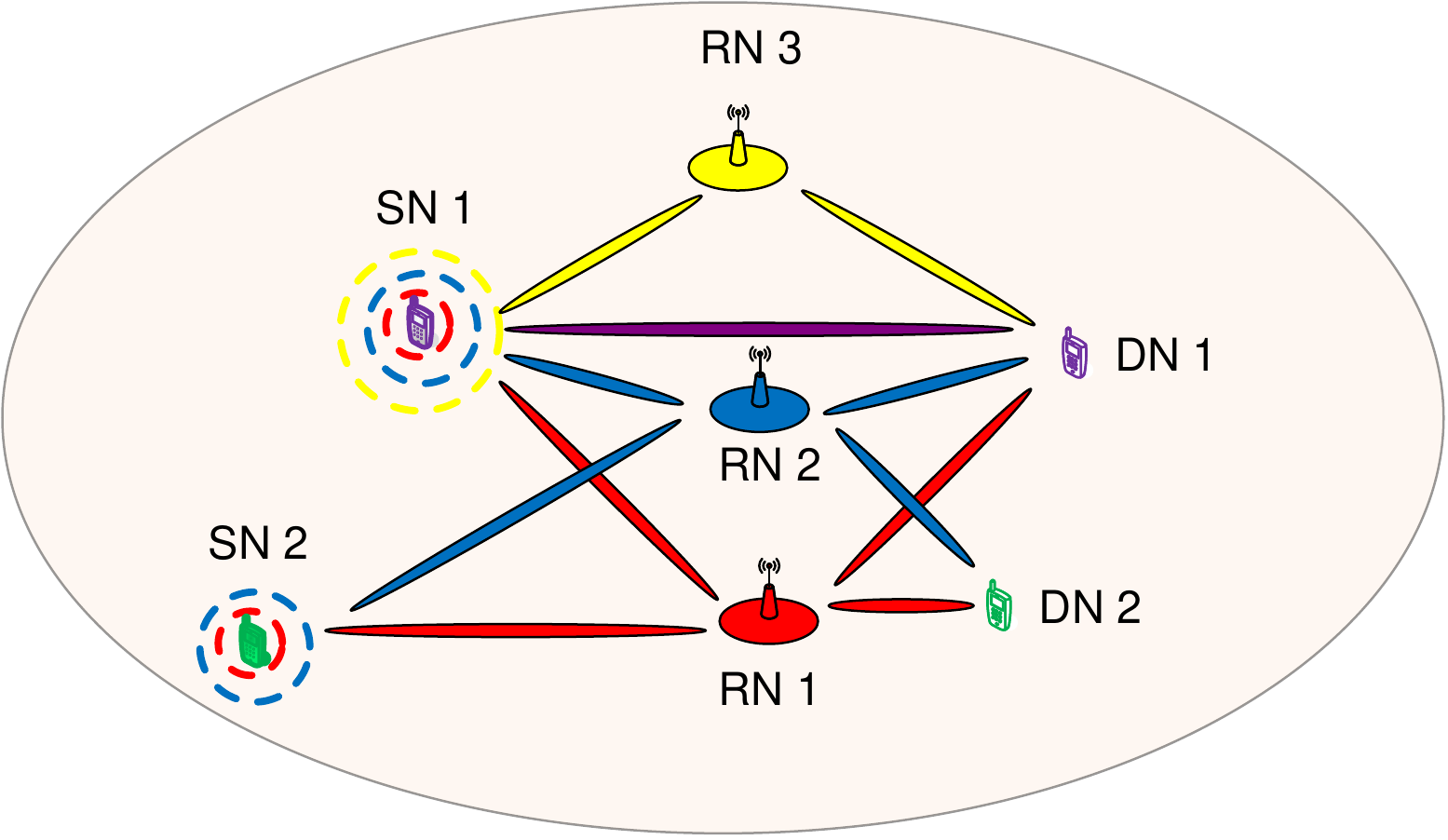}
\caption{Colored beams in SN indicate serving RN candidates}
\label{relay}
\end{subfigure}%
\hspace*{\fill}
\caption{Cooperative wireless system model.}\label{Coordinated}
\vspace{-1.5em}
\end{figure*}

The next-generation cellular networks, including cloud radio access networks (C-RAN) and software defined wireless networks (SDN), have proposed to enable cooperative transmission through the base band unit pool \cite{ran} and the controller \cite{Software}. One typical technology that has already been adopted by 3GPP Long Term Evolution is coordinated multi-point (CoMP) transmission \cite{3GPP}. As illustrated in Fig. \ref{cellular}, the adjacent outer cells and cell edge users can be considered as a new virtual cell (shaded area). This virtual cell is surrounded by multiple inner cells, has multiple base stations (BSs) serving as power sources, and works on the outer band (allocated by the fractional frequency reuse scheme \cite{freq}) or major subcarriers (allocated by the soft frequency reuse scheme \cite{freqboth}). The features of multiple power sources and shared spectrum have motivated the coordination transmission, which is considered as an effective tool to improve the coverage of high data rates and the cell-edge throughput.

In addition to the cellular networks with simultaneous multiple data transmissions, cooperative communications has been widely adopted in ad-hoc networks and cognitive networks \cite{6189412}, where cooperative sequential data transmissions play a major role. As illustrated in Fig. \ref{relay}, destination node (DN) combines the signal transmitted by source node (SN) in the first time slot and the forwarding signal transmitted by the relay node in the second time slot. Network coding based two-way relay schemes with decoding (decode-and-forward) and without decoding (amplify-and-forward, denoise-and-forward, compress-and-forward) are introduced to implement cooperative communications \cite{relayDef}. The inherent cooperative diversity can save energy by combining the signals received from different spatial paths and consecutive time slots \cite{ccDEF}.

Capitalizing on the internal flexibility of FDMA/OFDMA in power loading across the frequency channels/subcarriers, and the external flexibility in multiple nodes serving as power sources, the resource allocation scheme for cooperative wireless systems can dynamically assign the limited resources (spectrum and power) to deliver the best quality of service to customers at the lowest cost; that is, the available resources are allocated to the user who can best exploit the resources according to the current channel state information (CSI). This multi-node multi-UE diversity gained from dynamic resource allocation improves the performance of cooperative wireless systems. 

In cellular networks, a couple of research works on optimal power allocation \cite{RA}, \cite{RA2} have been conducted to maximize the system throughput with power constraints. Kivanc \emph{et al}. \cite{MA} investigated the subcarrier assignment problem to minimize the total power consumption while meeting data rate requirements. In ad-hoc networks and cognitive radio (CR) networks, to maximize the system capacity, Gong \emph{et al}. \cite {FDM} studied the optimal bandwidth and power allocation in a FDMA CR network. Chen and Wang \cite {Cognitive} presented a joint subcarrier and power allocation algorithm for multiuser OFDMA CR systems. Using the Lambert-W function, Brah \emph{et al}. \cite{Lambert} derived the closed form solution for subcarrier allocation in OFDMA based wireless mesh network. In wireless networks with hybrid energy supplies (an energy harvester and a constant energy source driven by a non-renewable resource), to minimize the on-grid energy consumption, Ng \emph{et al}. \cite{Hybrid} proposed the energy-efficient resource allocation for OFDMA systems. Han and Ansari \cite{hybridmulticell}, \cite{ICe} developed an energy aware cell size adaptation algorithm. However, the power source of each UE is limited to only one BS in the above works, i.e., each UE can only associate with one BS at any time.

In relay networks, for a given total power budget/time duration of the SN and RN, Mo \emph{et al}. \cite{6784109} optimized the power/time variables such that the outage probability of the cooperative relaying protocol is minimized. Assuming equal-time allocation for the SN and RN, Zhang \emph{et al}. \cite{6702843} investigated joint subcarrier and power allocation schemes to optimize the sum rate of downlink multi-UE transmission. For the system with node-specific power constraint, Luo \emph{et al}. \cite{Water} presented the joint water-filling (Jo-WF) power allocation algorithm to maximize the system throughput with two BSs jointly transmitting their constrained power to the multiple orthogonal subchannels.

Note that although various resource allocation algorithms have been extensively investigated to maximize the efficiency of cooperative wireless systems, most existing works, similar to the ones mentioned above, considered the scenario with only one or two serving nodes, and there are rather limited studies on the scenario with multiple serving nodes. Zhu and Wang \cite{6601771} studied the downlink throughput maximization problem for the multi-UE distributed antenna system. With a system-specific power budget, the maximal ratio transmission (MRT) scheme is adopted to allocate power among multiple distributed antennas, where each antenna can be considered as a BS. Sadek \emph{et al}. \cite{4034243} minimized the symbol error rate (SER) of the system with multiple sequential RNs, where each RN coherently combines the signals received from all of the previous RNs in addition to the signal from SN. They proposed a power allocation scheme with a system-specific budget on the total transmission power. Cui \emph{et al}. \cite{Water2} proposed the joint minimization power consumption (JMPC) algorithm to minimize the total power consumption subject to the system-specific throughput requirement and BS-specific power budget.


For the scenario with multiple cooperative nodes and node-specific power budget, we further extend the JMPC algorithm, and analyze the joint spectrum and power allocation problem in this paper. That is, using a new approach with ``power shifting'' and ``common candidate vector'', we aim to minimize the total transmission power consumption of the system while guaranteeing the UE-specific QoS requirement in terms of throughput.


The contributions of this paper include 1) we have proven the conjecture that for the system with arbitrary $M$ cooperative BSs and $N$ UEs, the minimum total power consumption can always be achieved when the number of \emph{multi-BS UEs} (UEs that are powered by multi-BS) is limited by $M-1$; 2) we have derived the UE-BS association scheme to determine the clusters of multi-BS UEs as well as the clusters of \emph{individual-BS UEs} (UEs that are powered by individual BSs); 3) we have proposed a complexity reduction scheme to improve the efficiency of the joint spectrum and power allocation algorithm (JSPA).

The major features that distinguish our work from the previous state-of-the-art works with similar system scenarios are summarized in Table \ref{comparision_t}. Since JSPA does not require UEs to have the capability to be served by all of the BSs, it is applicable to any cooperative networks where mobile UEs can move out of the coverage of certain BSs. Since JSPA is proven to be optimal, it outperforms the existing algorithms, and its low-complexity is desirable for the practical operation of the cooperative networks, such as the online resource allocation schemes which cope with mobile UEs.

\begin{table}[h]
\centering
  \caption{Comparison between JSPA and existing works}\label{comparision_t}
{\renewcommand{\arraystretch}{1.5}  
  \begin{tabular}{| c | c | c | c | c | c |}\hline
BS&	\multirow{2}{*}{Alg.}	&Maximum     &Serving BS 	&\multirow{2}{*}{OPT}&\multirow{2}{*}{CPX}		 \\
No.&                            &Multi-BS UE &Candidates    &                    &                            \\\hline
	
\multirow{2}{*}{2}& MRT	&$N$&	\multirow{2}{*}{2}	&{Sub-opt}&	Low\\\cline{2-3}\cline{5-6}
                  &JMPC	&\multirow{2}{*}{1}&  &\multirow{2}{*}{Opt}&High			\\\cline{2-2}\cline{4-4}\cline{6-6}
                  &JSPA &                  &Varying&	               &Low  	\\ \hline
\multirow{3}{*}{$M$}& MRT	&$N$&	\multirow{2}{*}{$M$}	&\multirow{2}{*}{Sub-opt}&	Low\\\cline{2-3}\cline{6-6}
	                &JMPC	            &1&			              &         &High\\\cline{2-6}
	                &JSPA	            &$M-1$&	Varying	&Opt&	Low\\ \hline
  \end{tabular}}
  \vspace{-1em}
\end{table}

 

\section{Coordinated Transmission Model}
Consider a cooperative downlink multi-user system, in which $M$ BSs coordinately assign spectrum and allocate power to $N$ users located in the coordinative zone, as depicted in Fig. \ref{cellular}. Each user feeds back the instantaneous CSI to its corresponding BS via a feedback channel. Through the backhaul channels, which can be optical fiber or out of band microwave links, each BS has access to the data and control information (such as CSI) of all of the $N$ users \cite{scen}. So, the data transmission of the users are dynamically coordinated among the BSs.

To simplify the mathematical derivation, we assume each BS has the same power constraint ${P_0}$ and share the same overall bandwidth $B_0$. Since multiple access technologies based on frequency division allocate orthogonal spectrum among UEs to avoid interference, $B_0$ is divided into $N$ distinct and nonoverlapping flat fading channels with various bandwidths, one for each UE. 

Furthermore, if UE $j$ is a multi-BS UE, the serving BSs would allocate the same channel to this UE, so that without shifting frequency, UE $j$ can optimally receive its information from the assigned channel with maximal-ratio combining (MRC) \cite{5740907}, \cite{6362722}. Thus, the achievable throughout of the $j$-th user given by the AWGN Shannon Capacity (sum rate) is expressed as
\begin{equation}\label{traffic}
{R_j}= {B_j}{{\log }_2}\left(1 + \frac{{\sum\limits_{i = 1}^M {{P_{i,j}}{{\left| {{H_{i,j}}} \right|}^2}} }}{{{\sigma _j}^2}}\right)
\end{equation}
where $B_j$ is the bandwidth assigned to the $j$-th channel, ${\sigma _j}^2 = {N_0}{B_j}$ represents the power of additive white Gaussian noise at the $j$-th channel,  ${P_{i,j}}$ denotes the allocated transmission power from BS $i$ to the $j$-th channel, and ${H_{i,j}}$ denotes the corresponding channel gain between BS $i$ and the $j$-th channel.

The goal here is to minimize the total transmission power of the system while meeting each user's throughput requirement ${R_j^S}$ as well as each BS's power and spectrum constraints. Since the circuit energy consumption associated with data reception is generally modeled as a constant \cite{1532220}, and we have assumed no frequency shifting for each UE, so only the transmission power of all of the BSs is considered in this paper.
\begin{equation}\label{eq:1}
\begin{array}{l}
{P_{overall}} =\min\sum\limits_{i = 1}^M\sum\limits_{j = 1}^N {{P_{i,j}}}\\
\begin{array}{*{20}{l}}
s.t.&{R_j} = {R_j^S},\  j \in\mathscr{N}\\
&\sum\limits_{j = 1}^N {{P_{i,j}}}  \le {P_0},\  i \in\mathscr{M}\\
&\sum\limits_{j = 1}^N {{B_{j}}}  = {B_0} \\
\end{array}
\end{array}
\end{equation}
where $\mathscr{M}=\{1,\cdots,M\}$, $\mathscr{N}=\{1,\cdots,N\}$. If UE $j$ has moved out of the coverage of BS $i$, or severe channel attenuation occurs such that UE $j$ cannot be associated with BS $i$, then set ${H_{i,j}}=0$.

For the sake of mathematical abbreviation, we denote ${\gamma _{i,j}} = ({P_0}{\left| {{H_{i,j}}} \right|^2})/\left({N_0}{B_0}\right)$, ${x_{i,j}} = {P_{i,j}}/{P_0}$ and ${y_{j}} = {B_{j}}/{B_0}$. Note that ${\gamma _{i,j}}$ is the signal to noise ratio (SNR) associated with BS $i$ over the total bandwidth $B_0$ when the entire power ${P_0}$ is allocated to the $j$-th UE. ${x_{i,j}}$ and ${y_{i,j}}$ represent the power and bandwidth allocation ratio, and ${R_j^S}/{B_0}=R'_j$. Since the logarithm is monotonically increasing, the objective function (\ref{eq:1}) combined with the constraints can be described as follows:
\begin{equation}\label{objxy}
\begin{array}{l}
Z =  \mathop{\min }\limits_{\{{\bf{X}},{\bf{Y}}\}} \sum\limits_{i = 1}^M\sum\limits_{j = 1}^N {{x_{i,j}}}\\
\begin{array}{*{20}{l}}
s.t. & {\sum\limits_{i = 1}^M {\gamma _{i,j}x_{i,j}}}   = ({2^{{R'_j}/{y_j}}}-1){y_{j}}, \ j\in\mathscr{N}\\
&\sum\limits_{j = 1}^N {{x_{i,j}}}  \le 1,  \ i \in\mathscr{M}   \\
&\sum\limits_{j = 1}^N {{y_{j}}}  = 1  \\
 \end{array}
  \end{array}
\end{equation}
where ${{\bf{X}}} = \{ {x_{i,j}}\left| {i \in\mathscr{M}; j \in\mathscr{N}} \right.\}$, ${{\bf{Y}}} = \{ {y_{j}}\left| j\in\mathscr{N} \right.\}$, and $Z={P_{overall}}/{P_0}\le M$ represents the total power consumption ratio.

\section{Problem Analysis}
As discussed in \cite{Water2}, finding the global optimal solution of the power allocation problem is very complicated. Solving (\ref{objxy}) is even more challenging due to the non-convexity of the joint optimization of spectrum and power. In order to achieve the minimum power consumption, we first decouple the power allocation problem from the spectrum allocation problem.
\subsection{Power allocation scheme}
The main result of this paper is the number of multi-BS UEs in the optimal solution is limited by $M-1$, as stated in the following Lemma.

\emph{Lemma 1}: For any spectrum allocation scheme ${{\bf{Y}}}$, there exists an optimal power allocation with at least $(M - 1)(N - 1)$ elements of ${{\bf{X}}}$ being zero.

The proof of Lemma 1 is provided in Appendix A. The observation presented in Lemma 1 simplifies the joint spectrum and power allocation problem greatly, because 1) the power allocation is decoupled from the spectrum allocation, which enables versatile access technologies, such as FDMA or OFDMA system; 2) the number of BS-UE links in the system, i.e., the number of non-zero elements in ${\bf{X}}$, is limited within the rage of $[N,N+M-1]$. 

\emph{Remark 1:} Define the \emph{SNR ratio} between (BS $i$, UE $j$) link and (BS $i'$, UE $j$) link as
\begin{equation}\label{ratio}
\gamma^{j}_{i,i'}=\frac{\gamma_{i,j}}{\gamma_{i',j}}
\end{equation}
According to the \emph{power shifting argument} in Appendix A, if $\gamma^{j}_{i,i'}$ allows a feasible power shifting that will decrease the total power consumption, then the corresponding power allocation is not optimal. So, power shifting argument can be used as an initial assessment to determine whether UE $j$ should be associated with BS $i$, $i'$ or both, and we will elaborate this in the next section. 

\subsection{UE-BS association scheme}
To satisfy the QoS requirements, each user $j$ should be associated with at least one BS $i$ such that $x_{i,j}y_{j}>0$. Since the number of non-zero elements in ${\bf{X}}$ is limited (Lemma 1), the majority of UEs will be associated with one BS only. According to the channel conditions, we will address the UE-BS association problem such that the complexity of finding the zero elements is further decreased.

Suppose there is no power limit for each BS, to minimize the power consumption of the system, the intuitive association scheme for each UE is to find the BS with the best channel condition. With this scheme, UE $j$ will be powered by BS $i$ only, where 
\begin{equation}\label{asso}
i=\arg\mathop {\max} \limits_{k\in\{1,2,\cdots,M\}}\{\gamma_{k,j}\}
\end{equation}
So, UEs will be divided into $M$ clusters denoted by \emph{initial disjoint clusters} $\{{J^0_i}\left|i\in\mathscr{M}\right.\}$, where the $i$-th cluster, ${J^0_i}$, consists of UEs, which prefer to be powered by BS $i$. 

With the introduction of BS-specific power budget, BS $i$ may not be able to power all of the UEs in cluster $i$, and other BSs will provide power coordination. Let $J^{i'}_{i}$ be the cluster consisting of UEs that 1) belong to ${J^0_{i}}$, and 2) are powered by BS $i'$ (partially or being taken over completely). Then, \emph{new disjoint clusters} $\{{J_i}\left|i\in\mathscr{M}\right.\}$ will be formed, where $J_i$ consists of UEs that are powered by BS $i$ only. 

\emph{Remark 2:} Since $J^{i'}_{i}\neq\Phi$ implies there is power shortage in BS $i$, $J^{i}_{i'}=\Phi$. Let $\overline{J^{i'}_{i}}=J_i^0\backslash{J^{i'}_{i}}$ be the UE cluster which consists of UEs that are not taken over by BS $i'$. If $\bigcup_{i'\in{M\backslash{i}}}{\overline{J^{i'}_{i}}}\subset{J_i^0}$, then BS $i$ would not take UEs from any other BSs, i.e., $\bigcup_{i'\in{M\backslash{i}}}J^{i}_{i'}=\Phi$, so ${J_i}=\bigcap_{i'\in{M\backslash{i}}}{\overline{J^{i'}_{i}}}$.


Since the stability of clusters depends on the SNR ratio $\gamma^{j}_{i,i'}$ in (\ref{ratio}), a \emph{common candidate (CC) vector} $\pmb{J^{CC}}$ can be defined for the disjoint clusters
\[\begin{array}{*{20}{c}}
\pmb{J^{CC}}={[{j_{1,2}}},&{{j_{1,3}}},& \cdots &{{j_{1,M}}},\\
{}&{{j_{2,3}}},& \cdots &{{j_{2,M}}},\\
{}&{}& \ddots & \vdots \\
{}&{}&{}&{{j_{M - 1,M}}]}
\end{array}\] 
where each element ${j_{i,i'}}$ is the multi-BS UE candidate that is commonly powered by BS $i$ and $i'$.

For any ${i}\in\{1,\cdots,M-1\}$ and ${i'}\in\{i+1,\cdots,M\}$, we require ${j_{i,i'}}$ in $\pmb{J^{CC}}$ to satisfy the following inequality
\begin{equation}\label{ccc}
\min\limits_{j\in{J_{i}}}{\gamma^{j}_{{i,i'}}}\ge {\gamma^{j_{i,i'}}_{{i,i'}}}>\max\limits_{j\in{J_{i'}}}{\gamma^{j}_{{i,i'}}}
\end{equation}
where for the initial cluster $J_{i}^0$ and $J_{i'}^0$, the corresponding $j_{i,i'}$ satisfies ${\gamma^{j_{i,i'}}_{{i,i'}}}\ge 1>{\gamma^{1+j_{i,i'}}_{{i,i'}}}$.

\emph{Lemma 2}: To minimize the power consumption, the CC vector which satisfies (\ref{ccc}) always exists for the optimal clusters $\{{J_i}\left|i\in\mathscr{M}\right.\}$.

The proof of Lemma 2 is provided in Appendix B. A very important point to be noticed from Lemma 2 is that multi-BS UEs are all in $\pmb{J^{CC}}$ because at most one UE will be associated with both BS $i$ and BS $i'$.  So
\[|\text{UNI}(\pmb{J^{CC}})|\le{M-1}\]
where $\text{UNI}(\bullet)$ consists of unique elements in $\bullet$.

Let ${J^{mul}_i}$ be the multi-BS UE candidates that are simultaneously powered by BS $i$ and other BSs, Then, we will have 
\begin{equation}\label{multi}
{J^{mul}_i}=(\bigcup\nolimits_{i'=i+1}^{M}{j_{i,i'}}\bigcup\nolimits_{i'=1}^{i-1}{j_{i',i}})\backslash(\bigcup\nolimits_{i'=1}^{M}{J_{i'}})
\end{equation}
\begin{equation}\label{defn}
\left\{\begin{array}{*{20}{l}}
x_{i,j}=({2^{{R'_j}/{y_j}}}-1){y_{j}}/\gamma_{i,j},&j\in{J_i}\\
{x_{i,j}=0},&j\notin {J_{i}}\bigcup{J^{mul}_i}\\
\end{array}\right.
\end{equation}
As we can see, by revealing the relationship between $J_i$ and $\pmb{J^{CC}}$, Lemma 2 can further differentiate the non-zero and zero variables of ${\bf{X}}$.


\subsection{Complexity reduction scheme}
Instead of iteratively solving (\ref{objxy}) for every $\pmb{J^{CC}}$ that satisfies (\ref{ccc}), we try to find the possible optimal $\pmb{J^{CC}}$ by considering the model of a 2-BS system, which can be used as the reference for the more complicated cooperative system involving three or more BSs.

For the initial disjoint clusters, suppose we relax the power constraint for BS $i_1$ and set the power limit of the other BS $i_2$, $\{i_1,i_2\}=\{1,2\}$, (\ref{objxy}) becomes:
\begin{equation}\label{relaxed_constraint}
\begin{array}{l}
Z=\min \sum\limits_{j \in {J_1^0}} \frac{({2^{\frac{{R{'_j}}}{{{y_j}}}}} - 1){y_j}}{{{\gamma_{1,j}}}}+\sum\limits_{j \in {J_2^0}} \frac{({2^{\frac{{R{'_j}}}{{{y_j}}}}} - 1){y_j}}{{{\gamma_{2,j}}}} \\
\begin{array}{*{20}{l}}
s.t.&\sum\limits_{j \in {J_{i_1}^0}} {({2^{{{R{'_j}}}/{{{y_j}}}}} - 1)}{y_j}/{{{\gamma_{{i_1},j}}}}  \le +\infty\\
&\sum\limits_{j \in {J_{i_2}^0}} {({2^{{{R{'_j}}}/{{{y_j}}}}} - 1)}{y_j}/{{{\gamma_{{i_2},j}}}}  \le 1\\
&\sum\limits_{j = 1}^N {{y_j}}  = 1\\
\end{array}
\end{array}
\end{equation}
where ${x_{i,j}=0}$ if $j\notin {J_i^0}$, and ${x_{i,j}}$ is given in (\ref{defn}) if $j\in {J_{i}^0}$.

As we can see, (\ref{relaxed_constraint}) is convex over ${\bf{Y}}$. By Lagrange dual function, we can derive the closed form solution expressed in the Lambert-W function \cite{Lambert}, or utilize various algorithms designed for the convex problem to approach the optimal solution \cite{Boyd:2004:CO:993483}. 

\begin{figure}[!htb]
\centering
\vspace{-1em}
\includegraphics[width=3in]{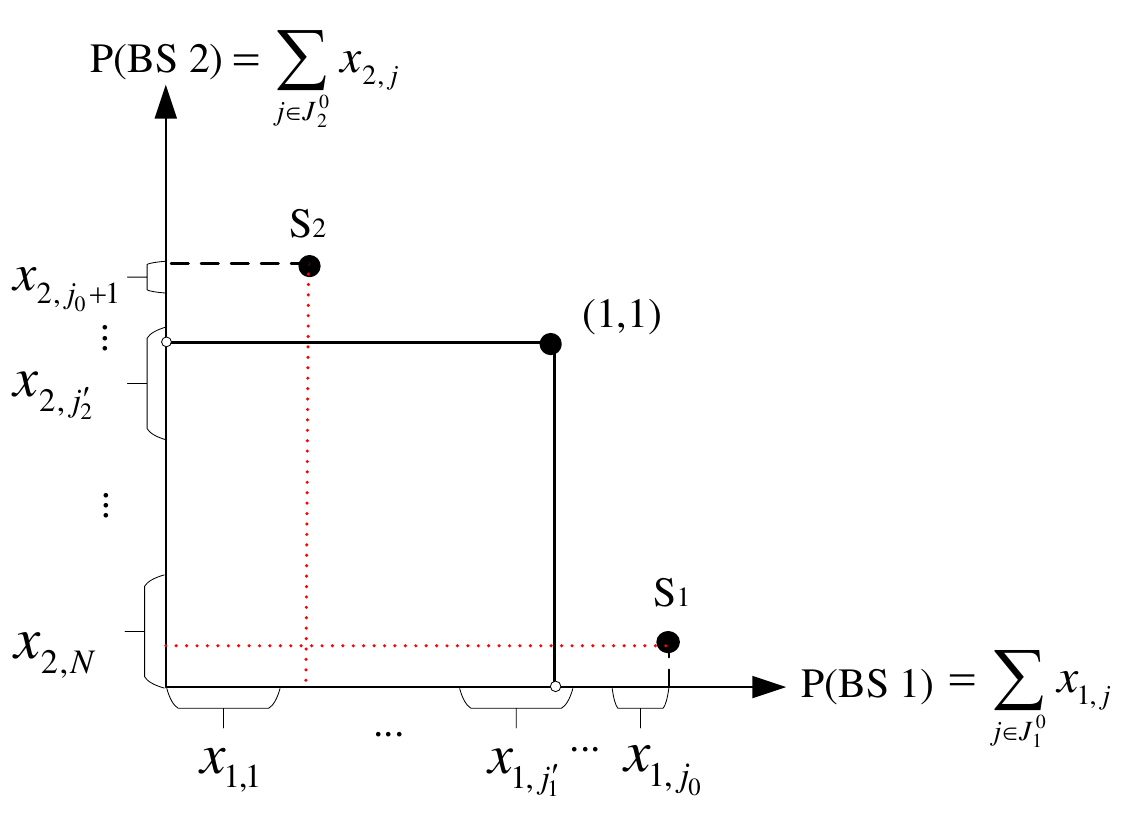}
\caption{Coordination with relaxed power constraint.}
\label{map}
\end{figure}

As shown in Fig. \ref{map}, the relaxed solutions can be represented in the two dimensional coordinates $S_i=(\sum\nolimits_{j \in {J_1^0}} x_{1,j},\sum\nolimits_{j \in {J_2^0}} {x_{2,j}})$, where $S_i$ is the solution to (\ref{relaxed_constraint}) with BS $i$ having no power budget. $S_i>({1,1})$ means the power consumption of BS $i$ is greater than 1, i.e., $S_i$ is located outside of the square region bound by $({1,1})$. $S_i\le({1,1})$ implies $S_i$ is located within the region bound by $({1,1})$.

\emph{Lemma 3}: Suppose UEs are sorted in the descending order of $\gamma_{1,2}^j$, $\pmb{J^{CC}}=\left[{j_{1,2}}\right]$.

1) If $S_1=S_2$, then $S_i$ is the optimal solution. 

2) If $S_{1}>({1,1})$, then there exists $j'_1\in J_{1}^0$ such that $\sum\nolimits_{j = 1}^{{j'_1-1}} {{x_{1,j}}}\le1$, $\sum\nolimits_{j = 1}^{{j'_1}} {{x_{1,j}}}>1$, then $j_{1,2}\ge j'_1$. 

3) If $S_{2}>({1,1})$, then there exists $j'_2\in J_{2}^0$ such that $\sum\nolimits_{j = j'_2+1}^{{N}} {{x_{2,j}}}\le1$, $\sum\nolimits_{j = j'_2}^{{N}} {{x_{2,j}}}>1$, then $j_{1,2}\le j'_2$.

The proof of Lemma 3 is provided in Appendix C. By relaxing the power constraint of a BS, Lemma 3 can limit the range of $j_{1,2}$, so that the complexity of finding the optimal solution is much lower than iterating through every possible $\pmb{J^{CC}}$.

\emph{Remark 3:} For any disjoint clusters $\{{J_i}\left|i\in\mathscr{M}\backslash\{i_1,i_2\}\right.\}$ with $M\ge3$, we can limit the range of $j_{i_1,i_2}$ by using Lemma 3. The details of how to apply Lemma 3 are provided in the next section.
\begin{figure}[!b]
\centering
\includegraphics[width=3in]{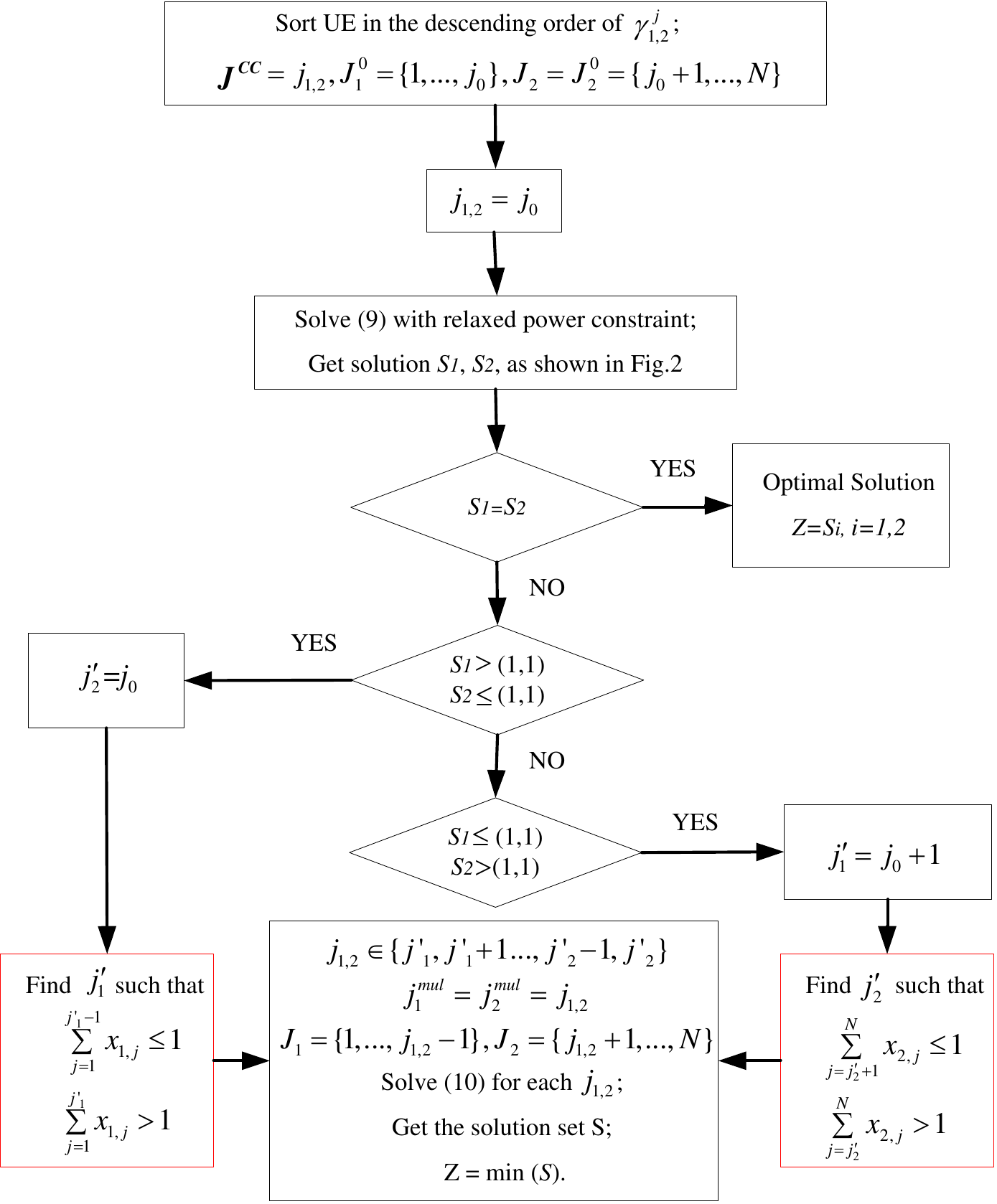}
\caption{JSPA algorithm, $M=2$.}
\label{algorithm}
\end{figure}
\section{Joint spectrum and power allocation algorithm}
Since $J^0_i\subseteq{J_i}$ implies BS $i$ may provide power coordination to other BSs, $M-\sum\nolimits_{i=1}^M1{\{J_i^0\subseteq{J_i}\}}$ BSs will receive power coordination, where $1\{\bullet\}=0$ if $\bullet$ is false and 1 otherwise. So, for any $\{{J_{i}},{J^{mul}_i}| i\in\mathscr{M}\}$ which satisfies Lemma 2, the objective function in (\ref{objxy}) can be transformed into
\begin{equation}\label{caseeq}
\begin{array}{l}
\mathop{\min }\sum\limits_{i=1}^M(\sum\limits_{j \in {J_i}}\frac{({2^{\frac{{R{'_j}}}{{{y_j}}}}} ){y_j}}{{{\gamma_{i,j}}}}+\sum\limits_{j\in{{J^{mul}_i}}}{x_{i,j}})1{\{J^0_i\subseteq{J_i}\}}+ M-\sum\limits_{i=1}^M1{\{J^0_i\subseteq{J_i}\}}\\
\begin{array}{*{20}{l}}
s.t. &\sum\limits_{j \in {J_i}} \frac{({2^{\frac{{R{'_j}}}{{{y_j}}}}} - 1){{{y_j}}}}{{{\gamma_{i,j}}}}+\sum\limits_{j\in{{J^{mul}_i}}}{x_{i,j}}\le 1,\ i\in\mathscr{M}\\
&{\sum\limits_{i = 1}^M {\gamma _{i,j}x_{i,j}}}   = ({2^{{R'_j}/{y_j}}}-1){y_{j}},\ j\in\bigcup\limits_{i=1}^{M}{J^{mul}_i}\\
& \sum\limits_{j = 1}^N {{y_j}}  = 1\\
\end{array}
\end{array}
\end{equation}
where ${x_{i,j}}$ is given in (\ref{defn}), and if $1\{J_i^0\subseteq{J_i}\}=0$, the first constraint is satisfied with equality, i.e., $\sum\limits_{j \in {J_i}} ({2^{{{R{'_j}}}/{{{y_j}}}}} - 1){{{{y_j}}}/{{{\gamma_{i,j}}}}}+\sum\limits_{j\in{{J^{mul}_i}}}{x_{i,j}}=1$.

Based on Lemmas 1-3, we first present the joint spectrum and power allocation (JSPA) algorithm with $M=2$. Similar to (\ref{relaxed_constraint}), for each $j_{1,2}$ in Fig. \ref{algorithm}, (\ref{caseeq}) is also a convex optimization problem with $N+2$ variables $\{{\bf{Y}},x_{1,j_{1,2}},x_{2,j_{1,2}}\}$. With arbitrary $M$, the two procedures to achieve optimal resource allocation are given in Alg. \ref{euclid} and Alg. \ref{euclid1}.
\begin{algorithm}
\caption{JSPA algorithm: UE-BS association}\label{euclid}
\begin{algorithmic}[1]
\For {$\pmb{J^{CC}}$ with $|\text{UNI}(\pmb{J^{CC}})|\le{M-1}$}
\LState $\mathscr{M}=\{1,\cdots,M\}$, $\mathscr{N}=\{1,\cdots,N\}$
    \LState According to (\ref{asso}), get ${J^{0}_{i}}$, $i\in\mathscr{M}$, $j\in\mathscr{N}$
    
    \LState $\overline{J_{i}^{i'}}={J_i^0}$, $i\in\mathscr{M}$, $i'\in\mathscr{M}\backslash{i}$
    \LState ${J_i},{J^{mul}_i}=\Phi$, $i\in\mathscr{M}$
    
    \While {$\bigcup_{i\in\mathscr{M}}{J_i}\bigcup_{i\in\mathscr{M}}{J^{mul}_i}\ne\mathscr{N}$}
     \LState $\mathscr{N}\leftarrow\mathscr{N}\backslash\bigcup_{i\in\mathscr{M}}{J_i}$
     \LState $\mathscr{M}\leftarrow\mathscr{M}\backslash\{i\in\mathscr{M}:\bigcup_{i'\in{\mathscr{M}\backslash{i}}}\overline{J_{i}^{i'}}\subset{J_i^0}\}$
     \LState Update $J_{i}^0$, $i\in\mathscr{M}$, $j\in\mathscr{N}$
     \For {$i\in\mathscr{M}$}
     \LState  $\mathscr{M'}=\{1,\cdots,M\}\backslash{i}$
     \LState $\overline{J_{i}^{i'}}=\{j\in{J^{0}_{i}}:\gamma^j_{i,i'}>\gamma^{j_{{i},i'}}_{i,i'}\}$, $i'\in\mathscr{M'}$
    \LState $J_{i}\leftarrow{J_{i}}\bigcup\{\bigcap_{i\in\mathscr{M'}}{\overline{J_{i}^{i'}}}\}$
     \EndFor
     \LState According to (\ref{multi}), get ${J^{mul}_i}$, $i\in\mathscr{M}$, $j\in\mathscr{N}$
     \EndWhile
     \LState\Return $\{\pmb{J^{CC}},{J_{i}},{J^{mul}_i}| i\in\{1,\cdots,M\}\}$  
\EndFor
\end{algorithmic}
\end{algorithm}

\begin{algorithm}
\caption{JSPA algorithm: Complexity reduction}\label{euclid1}
\begin{algorithmic}[1]
\LState $Z=M$
\For {$i_1,i_2\in\{1,\cdots,M\}$, ${i_2>i_1}$}
\LState {$\mathscr{M}\leftarrow\{1,\cdots,M\}\backslash\{i_1,i_2\}$}
\For {$\{{J_{i}},{J^{mul}_i}| i\in\mathscr{M}\}$ returned by Alg. \ref{euclid}}
\LState {$\mathscr{N}\leftarrow\{1,\cdots,N\}\backslash\{\bigcup_{i\in\mathscr{M}}{J_i}\bigcup_{i\in\mathscr{M}}{J^{mul}_i}\}$}
\LState Sort UE in the descending order of $\gamma_{i_1,i_2}^j$
\LState Update ${J^{0}_{i}}$, $i\in\{i_1,i_2\}$, $j\in\mathscr{N}$
\LState In (\ref{caseeq}), ${J_{i}}\leftarrow{J^{0}_{i}}$, $i\in\{i_1,i_2\}$
\LState Keep power budgets of BSs other than $i_1/i_2$
\LState Get the relaxed solutions to (\ref{caseeq}): $S_{i_1}/S_{i_2}$

\If {$S_{i_1}>(1,1)$, $S_{i_2}\le(1,1)$}
\LState Get $j'_{i_1}$ according to Lemma 3
\LState $j'_{i_2}=\max\nolimits_{j\in{J_{i_1}^0}}\{j\}$

\ElsIf {$S_{i_2}>(1,1)$, $S_{i_1}\le(1,1)$}
\LState Get $j'_{i_2}$ according to Lemma 3
\LState $j'_{i_1}=\max\nolimits_{j\in{J_{i_1}^0}}\{j\}+1$

\ElsIf {$S_{i_1}>(1,1)$, $S_{i_2}>(1,1)$}
\LState Get $j'_{i_1}$, $j'_{i_2}$ according to Lemma 3

\ElsIf {$S_{i_1}\le(1,1)$, $S_{i_2}\le(1,1)$}
\LState $j'_{i_1}=j'_{i_2}=\max\nolimits_{j\in{J_{i_1}^0}}\{j\}$

\EndIf

\For {$\{{J_{i}},{J^{mul}_i}|i\in\{i_1,i_2\}\}$ returned by Alg. \ref{euclid}}
\If {$j_{i_1,i_2}\in\{j'_{i_1},\cdots,j'_{i_2}\}$}
\LState {Get the solution to (\ref{caseeq}): $S$}
\LState $Z\leftarrow\min\{Z,S\}$
\EndIf
\EndFor

\EndFor
\EndFor
\LState\Return $Z$
\end{algorithmic}
\end{algorithm}

\section{Simulation Results}
We assume that 20 independent and identically distributed (i.i.d.) Rayleigh-faded users are uniformly located within the shaded zone (see Fig. \ref{Coordinated}). 
$R$ is 1000 m and the inner cell radius $R'$ is 600 m. The distance-dependent path loss model is $L(d) = 128.1 + 37.6\lg (d)$ dB, $d$ in km, and ${N_0} =  -174$ dBm/Hz. For the sake of simplicity, we assume $B_0= 1$, and each BS's power constraint is $P_0=1$.

The performances of the proposed JSPA algorithm and JMPC algorithm in \cite{Water2} are averaged over 1,000 independent snapshots by \emph{Monte-Carlo} simulation. The throughput requirement of each UE is defined as 
\begin{equation}\label{simu}
{R_j}=\epsilon{R^0_j}\text{, }\epsilon>0
\end{equation}
where ${R^0_j}$ is generated according to (\ref{traffic}), with all of the $N$ users being assigned equal spectrum and power (ESP) from each BS.

\subsection{Two-node system}
\begin{figure}[!htb]
\centering
\includegraphics[width=3in]{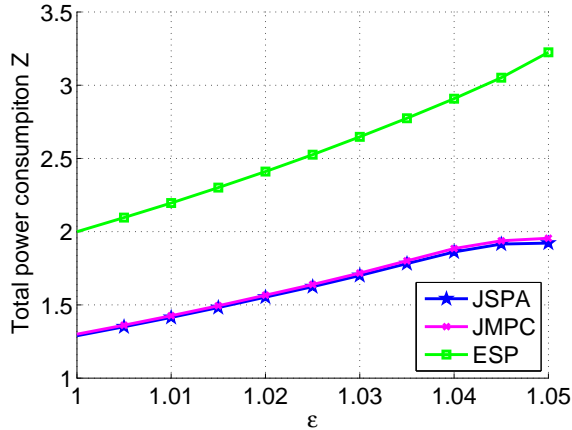}
\caption{Total power consumption of the 2-BS cooperative system.}\label{fig2}
\end{figure}

As pointed out in Table \ref{comparision_t}, for $M=2$, both JSPA and JMPC are optimal in the sense of power allocation. For easy comparison, we assume the system-specific throughput requirement of JMPC is divided among all of the UEs, i.e., UE-specific $R_j^S$ in (\ref{simu}). Then, as we can see in Fig.\ref{fig2}, JSPA outperforms JMPC in the total power consumption. The reason is that JSPA supports flexible spectrum allocation, while JMPC adopts equal bandwidth allocation for all UEs.  

\begin{figure}[!htb]
\centering
\includegraphics[width=3in]{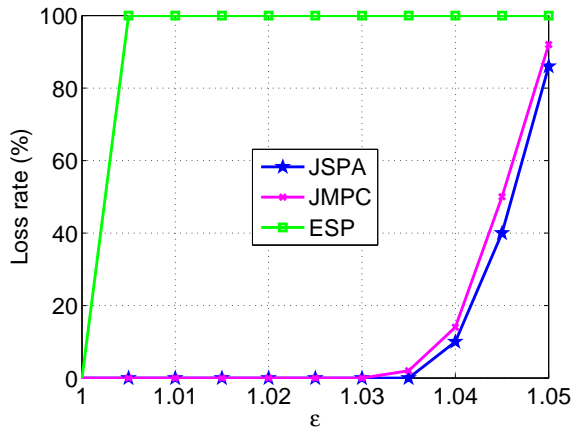}
\caption{Loss rate of the 2-BS cooperative system.}\label{LOSS}
\end{figure}

Fig. \ref{LOSS} indicates that with $\epsilon>1$, there will be loss, i.e., the system fails to support all of the $N$ users' throughput requirements with its maximum power and spectrum resources. Apparently, the loss rate must be zero with $\epsilon \le 1$ and the loss rate will increase with $\epsilon$. Since JSPA always consumes less or equal power, the loss rate is smaller accordingly. 

\subsection{Multi-node system}
To verify the point that for $M\ge3$, JSPA is optimal while JMPC is sub-optimal in the sense of power allocation, we assume instead of UE-specific throughput requirement and spectrum allocation, JSPA requires UE-common bandwidth allocation and system-specific throughput requirement (sum of ${R_j}$ in (\ref{simu})).

\begin{figure}[!htb]
\centering
\includegraphics[width=3in]{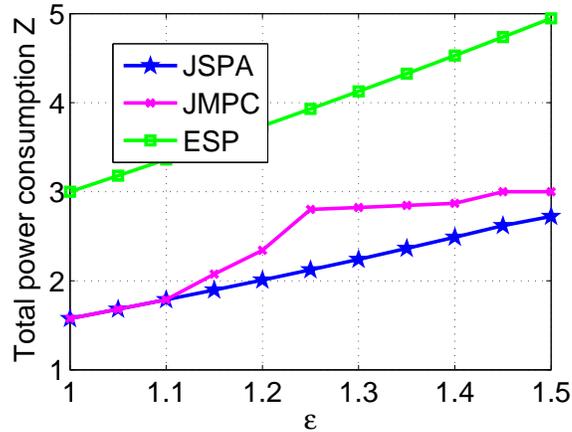}
\caption{Total power consumption of the 3-BS cooperative system.}\label{fig3}
\end{figure}

\begin{figure}[!htb]
\centering
\includegraphics[width=3in]{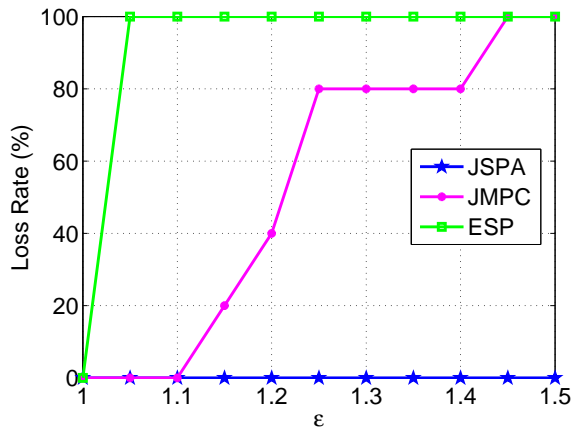}
\caption{Loss rate of the 3-BS cooperative system.}\label{LOSS2}
\end{figure}

As we can see in Figs. \ref{fig3} and \ref{LOSS2}, JSPA always achieves the best performance, both in total power consumption and the loss rate, even when it does not enable flexible spectrum allocation. The gain will increase with the randomness of $R_j$ rather than with the simultaneous increase/decrease with bigger/smaller $\epsilon$. The randomness make it less likely for one BS to exceed the power limit, when it provides spectrum or power to help the overloaded BS. Since JSPA and JMPC are compared in the same enviorment, we can conclude that under the perfect coordinated transmission between the multiple BSs, the proposed JSPA algorithm provides a significant reduction in the power consumption.
\section{Conclusion}
In this paper, we have investigated the joint spectrum and power allocation problem in minimizing the overall transmit power consumption while meeting the throughput requirements of each UE and each BS's power constraint for a cooperative downlink multi-user system. We have shown analytically that the number of multi-BS UE should be limited by the number of BSs. Moreover, we have also proposed the UE-BS association scheme and the corresponding complexity reduction scheme, which determines the serving BSs for each UE based on channel conditions and the constraints in the optimization problem. Finally, a novel joint spectrum and power allocation algorithm, proven to yield the minimum total power consumption, is proposed. Although the system model is based on downlink cellular network, the derived results are applicable for various networks with cooperative features: multiple power sources and shared spectrum.


%

\appendices
\section{Proof of Lemma 1}\label{a}
First, we will prove that Lemma 1 is true for $M=2$. Then, mathematical induction is used to prove the scenario for $M\ge3$.

\subsection{$M=2$}
Since all of the $N$ users need power, either from BS 1, BS 2, or both, then at least $N$ elements of ${{\bf{X}}}$ are non-zero. To prove Lemma 1, we need to prove that the minimum power consumption can be guaranteed when at most one of the $N$ users is served by two BSs simultaneously.

\begin{figure}[!htb]
\vspace{-1em}
\centering
\includegraphics[width=2.5in]{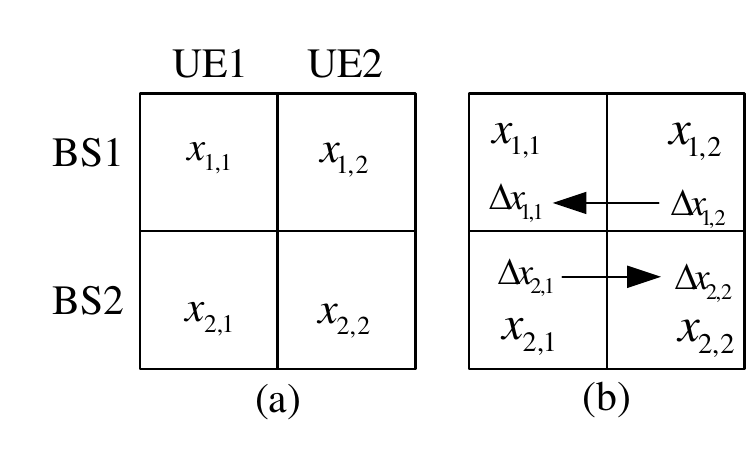}
\vspace{-1em}
\caption{Power shift in the 2-BS cooperative wireless system.}
\label{Power}
\end{figure}

We use reductio ad absurdum here. Suppose in the optimal solution $\{{{\bf{X}}},{{\bf{Y}}}\}$, both UE 1 and UE 2 are powered by BS 1 and BS 2, i.e., ${x_{i,j}} > 0$, $\left(i,j = 1,2\right)$. Then, any power shift $\Delta {x_{i,j}}\neq0$, $\left( i,j = 1,2\right)$ in Fig. \ref{Power} (b) will result in higher total power consumption.
\[\begin{array}{*{20}{c}}
{{{x'}_{1,1}} = {x_{1,1}} + \Delta {x_{1,1}}},{{{x'}_{1,2}} = {x_{1,2}} - \Delta {x_{1,2}}}\\
{{{x'}_{2,1}} = {x_{2,1}} - \Delta {x_{2,1}}},{{{x'}_{2,2}} = {x_{2,2}} + \Delta {x_{2,2}}}
\end{array}\]

To guarantee the throughput requirement, ${\sum\nolimits_{i = 1}^M {\gamma _{i,j}x_{i,j}}}$ should remain the same after the power shift, which yields
\[\begin{array}{l}
{\gamma_{1,1}}\Delta {x_{1,1}} - {\gamma_{2,1}}\Delta {x_{2,1}} = 0\\
{\gamma_{2,2}}\Delta {x_{2,2}} - {\gamma_{1,2}}\Delta {x_{1,2}} = 0
\end{array}\]

If ${\gamma_{1,1}}{\gamma_{2,2}} - {\gamma_{1,2}}{\gamma_{2,1}} \ge 0$, there always exists a power shift to ensure
$\Delta{x_{1,2}} + \Delta{x_{2,1}} -\Delta{x_{1,1}}- \Delta{x_{2,2}}\ge 0$, where
\[\left\{ \begin{array}{l}
\Delta {x_{2,1}} \ge \Delta {x_{2,2}} \ge 0,\Delta {x_{1,1}} = \Delta {x_{1,2}} \ge 0\\
\Delta {x_{1,2}} \ge \Delta {x_{1,1}} \ge 0,\Delta {x_{2,1}} = \Delta {x_{2,2}} \ge 0
\end{array} \right.\]

Similarly, for ${\gamma_{1,1}}{\gamma_{2,2}} - {\gamma_{1,2}}{\gamma_{2,1}} \le 0$, the following $\Delta {x_{i,j}}$ is always feasible such that the total power consumption will decrease after the power shifting.
\[\left\{ \begin{array}{l}
\Delta {x_{2,2}} \le \Delta {x_{2,1}} \le 0,\Delta {x_{1,1}} = \Delta {x_{1,2}} \le 0\\
\Delta {x_{1,1}} \le \Delta {x_{1,2}} \le 0,\Delta {x_{2,1}} = \Delta {x_{2,2}} \le 0
\end{array} \right.\]
This contradicts with the assumption. 
 
If there is a set consisting of more than two users powered by BS 1 and 2 simultaneously, we can iteratively group these users into pairs, and do power shifting as in the two-user case. The remaining users that are powered by two BSs form a new user set. Finally, we will get at most one user being served by the two BSs simultaneously.

\emph{Note 1}: 1) For ${\gamma_{1,1}}{\gamma_{2,2}} - {\gamma_{1,2}}{\gamma_{2,1}} < 0$, it is always more power efficient if BS 1 schedules more power to UE 2, while UE 1 prefers BS 2; the optimal power allocation must be the case when at least one of $x'_{1,1}$ and $x'_{2,2}$ is zero. 
2) For ${\gamma_{1,1}}{\gamma_{2,2}} - {\gamma_{1,2}}{\gamma_{2,1}} > 0$, the optimal power allocation must be the case when at least one of $x'_{1,2}$ and $x'_{2,1}$ is zero. 3) For ${\gamma_{1,1}}{\gamma_{2,2}} - {\gamma_{1,2}}{\gamma_{2,1}} = 0$, if the power keeps shifting until at least one of $x_{i,j}$ $(i,j=1,2)$ becomes zero, it can still guarantee the minimum power consumption because power shifting brings no increment in the total power consumption.

\subsection{$M\ge3$}
Assume Lemma 1 is true for the scenario with $M-1$ BSs.

\begin{figure}[!htb]
\centering
\includegraphics[width=3.2in]{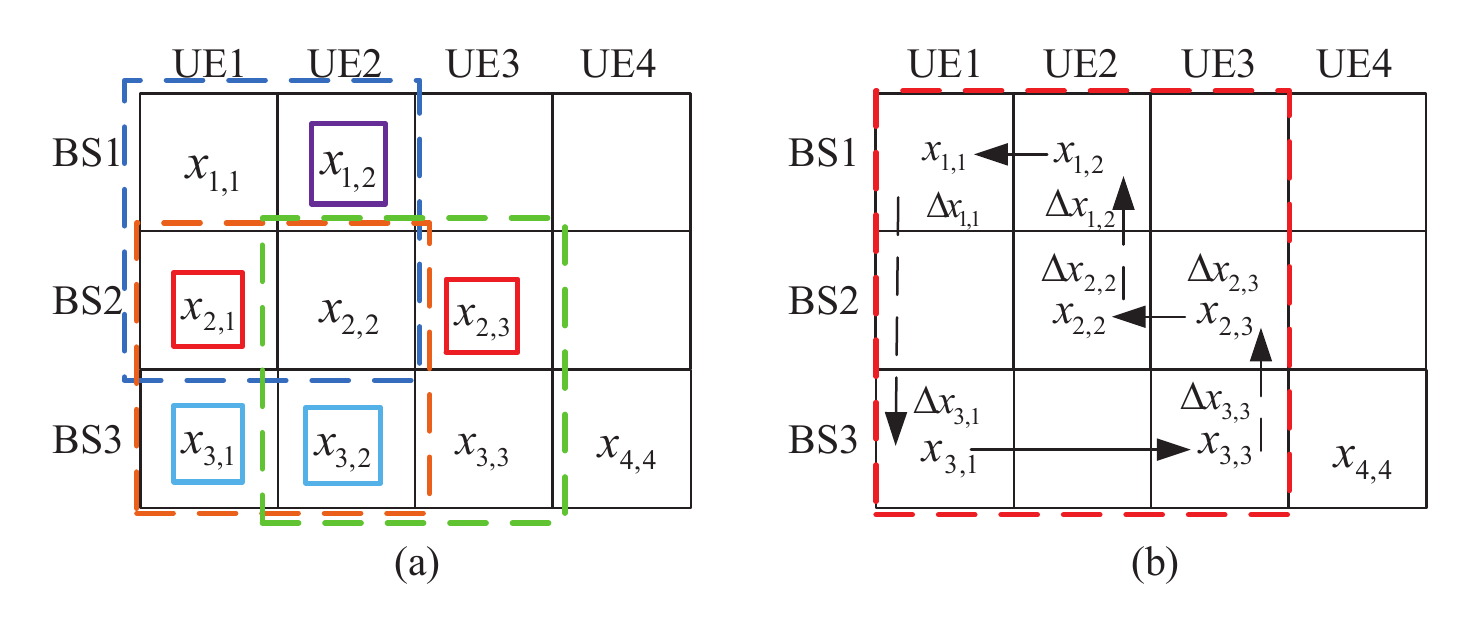}
\vspace{-1em}
\caption{Power shift in the 3-BS cooperative wireless System.}
\label{OFDM}
\end{figure}

For the scenario with $M$ BSs and $N$ UEs, we represent the power allocation solution as an $M\times{N}$ area, similar to the ${3}\times{4}$ area in Fig. \ref{OFDM} (a). 

Suppose there are more than $MN-(M-1)(N-1)=M+N-1$ non-zero cells in the $M\times{N}$ area, taking $M+N$ non-zero cells for example. Since each UE, i.e., column, has at least one non-zero cell, then at least $N-M$ UEs will be powered by individual BSs. 

Suppose columns $\{M+1,\cdots,N\}$ of the $M\times{N}$ area represent UEs that are powered by individual BSs, then there must be $2M$ non-zero cells in the $M\times{M}$ square formed by the first $M$ columns of the $M\times{N}$ area (similar to the first $3\times {3}$ square in Fig. \ref{OFDM} (b)). 

Denote $T_i$, $i\in\{1,\cdots,M\}$ as the number of non-zeros elements in the $i$-th row of the $M\times{M}$ square. Since $\sum\nolimits_{i=1}^{M}{T_i}\le2M$ and Lemma 1 is true for $M-1$ BSs, then in the $M\times{M}$ square, we can get
\[2M-T_i\le{\left(M-1\right)+\left(M-1\right)},\forall i \in\{1,\cdots,M\}\]

The above equation implies $T_i=2, \forall i\in\{1,\cdots,M\}$, i.e., each row has two non-zero cells in the $M\times{M}$ square. If there are columns that have only one non-zero cell, delete these columns and the corresponding rows where these non-zero cells are located. Finally, an $m\times{m}$, $(M\ge{m}\ge2)$ sub-square which has two non-zero cells in each column and each row must exist. Divide the $2m$ non-zeros cells into two groups, each with $m$ non-zero elements. Within each group, each row is distinct and so is each column.

Taking the $3\times 3$ dotted sub-square in Fig. \ref{OFDM} (b) for example. Group 1 includes $\{x_{1,1},x_{2,2},x_{3,3}\}$ and group 2 includes $\{x_{3,1},x_{1,2},x_{2,3}\}$. Suppose the optimal solution includes the $2m$ non-zero elements corresponding to the non-zero cells in the $m\times{m}$ sub-square. Then, the total power consumption must increase with the new power allocation $\{x'_{i,j}\}$ after the power shift $\{\Delta {x_{i,j}}\}$. 

\[\begin{array}{l}
\begin{array}{*{20}{c}}
{{{x'}_{1,1}} = {x_{1,1}} + \Delta {x_{1,1}}},{{{x'}_{1,2}} = {x_{1,2}} - \Delta {x_{1,2}}}\\
{{{x'}_{2,2}} = {x_{2,2}} + \Delta {x_{2,2}}},{{{x'}_{2,3}} = {x_{2,3}} - \Delta {x_{2,3}}}
\end{array}\\
\begin{array}{*{20}{c}}
{{{x'}_{3,1}} = {x_{3,1}} - \Delta {x_{3,1}}},{{{x'}_{3,3}} = {x_{3,3}} + \Delta {x_{3,3}}}
\end{array}
\end{array}\]
In order to guarantee the throughput requirement of each user, the power that shifted out should be equal to the amount that shifted in, so that the received power remains the same.
\[\begin{array}{l}
{\gamma_{1,1}}\Delta {x_{1,1}} = {\gamma_{3,1}}\Delta {x_{3,1}}\\
{\gamma_{2,2}}\Delta {x_{2,2}} = {\gamma_{1,2}}\Delta {x_{1,2}}\\
{\gamma_{3,3}}\Delta {x_{3,3}} = {\gamma_{2,3}}\Delta {x_{2,3}}
\end{array}\]
Hence, for ${\gamma_{1,1}}{\gamma_{2,2}}{\gamma_{3,3}} \le {\gamma_{1,2}}{\gamma_{3,1}}{\gamma_{2,3}}$, the power shift that satisfies the following condition will bring no increment to the total power consumption: $\Delta {x_{2,2}} \le \Delta {x_{2,3}} \le 0$, $\Delta {x_{1,1}} = \Delta {x_{1,2}} \le 0$, $\Delta {x_{3,3}} = \Delta {x_{3,1}} \le 0$.

For ${\gamma_{1,1}}{\gamma_{2,2}}{\gamma_{3,3}} > {\gamma_{1,2}}{\gamma_{3,1}}{\gamma_{2,3}}$, when $\Delta {x_{2,3}} > \Delta {x_{2,2}} > 0,\Delta {x_{1,1}} = \Delta {x_{1,2}} > 0,\Delta {x_{3,3}} = \Delta {x_{3,1}} > 0$, the total power consumption can be further decreased. This contradicts with the assumption that any power shift will increase the total power consumption. 

The power shift can continue until at least one of the $\{ {x_{1,1}},{x_{1,2}},{x_{2,2}},{x_{2,3}},{x_{3,1}},{x_{3,3}}\}$ becomes zero, and the number of non-zero elements in the $3\times3$ sub-square should be no greater than $2\times3-1$.

In summary, if the product of $\gamma_{i,j}$ of cells in group 1 is greater than the product of SNR of group 2, then a power shift from group 2 to group 1 that will decrease the total power consumption exists. If the product of group 1 is no more than the group 2's product, then there exists a power shift from group 1 to group 2 that will bring no increment to the total power consumption. 

In either case, the power shifting will continue until at least one of the cells becomes zero. So, the number of non-zero cells in the $m\times{m}$ square will be no more than $2M-1$. The number of non-zero cells in the $M\times {N}$ area will be no more than $(2M-1)+(N-M)=M+N-1$. Accordingly, the number of zeros cells will be greater than $MN-(N+M-1)=(M-1)(N-1)$.

When there are more than $M+N$ non-zero cells in the $M\times{N}$ area, we can always iteratively take $M+N$ non-zero cells and do power shifting to make the number of the non-zero cells no more than $M+N-1$.

\section{Proof of Lemma 2}\label{b}
We will prove that Lemma 2 is true for the scenarios for $M=2,3$. For scenarios with $M>3$, the same argument can be utilized to arrive at the same conclusion.
\subsection{$M=2$, $\pmb{J^{CC}}=\left[{j_{1,2}}\right]$}

To prove Lemma 2, we need to show in the optimal clusters $J_{1}$ and $J_{2}$, when users are sorted in descending order of ${\gamma^j_{1,2}}$, 
\begin{equation}\label{con}
\max\limits_{j\in{J_{1}}}\{j\}<\min\limits_{j\in{J_{2}}}\{j\}
\end{equation}

For the initial clusters $J_1^0=\{1,\cdots,j_0\}$ and $J_2^0=\{j_0+1,\cdots,N\}$, let $j_{1,2}=j_0$. 

For the optimal clusters, suppose there are two users $j_1^* \in {J_1},j_2^* \in {J_2}$, and $j_1^*>j_2^*$, as shown in Fig. \ref{association}. Since ${\gamma_{1,j_1^*}}{\gamma_{2,j_2^*}} - {\gamma_{2,j_1^*}}{\gamma_{1,j_2^*}} < 0$, power shifting between the two users will result in only one of the following scenarios (\emph{Note 1}):

\begin{figure}[!htb]
\centering
\vspace{-1em}
\includegraphics[width=3in]{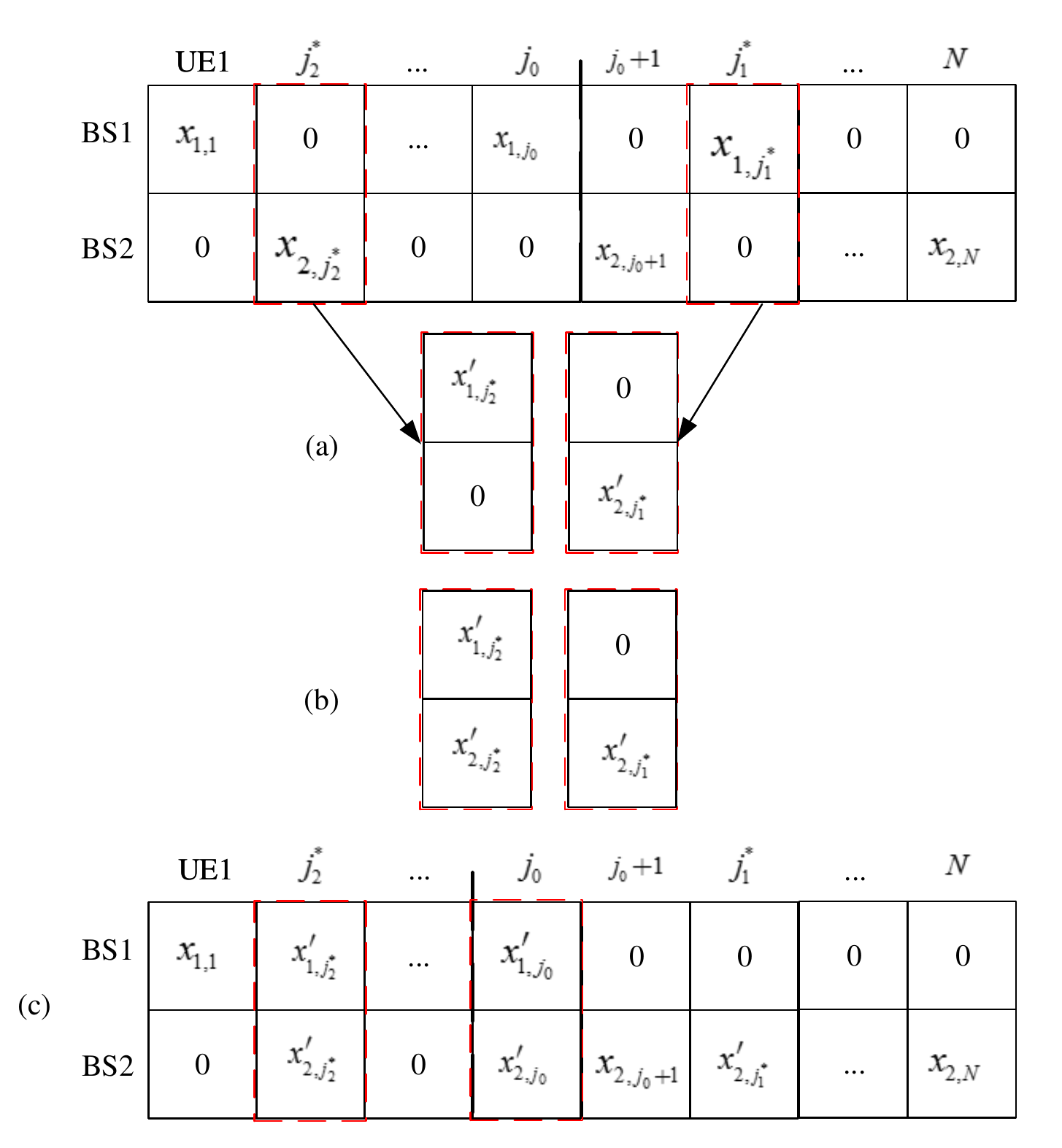}
\caption{UE-BS association in the 2-BS cooperative wireless system.}
\label{association}
\end{figure}

\emph{Scenario 1}: As shown in Fig. \ref{association} (a), ${J_1} = {J^0_1}$, ${J_2} ={J^0_2}$. 

\emph{Scenario 2}: One of the users is powered by two base stations; take $j_2^*$ for example, as shown in Fig. \ref{association} (b). Then, the power shifting between UE $j_2^*$ and $j_0$ will result in Fig. \ref{association} (c), where $x'_{2,j_2^*}x'_{1,j_0}=0$. 

1) If $x'_{2,j_2^*}=0$, we will have ${J_1}=\{1,\cdots,j_0-1\}$, $\{j_0+1,\cdots,N\}\subset{J_2}$. 

2) If $x'_{2,j_2^*}>0$, $x'_{1,j_0}=0$, then UE $j_2^*$ will continue to do power shifting with UE $\{j_0-1,j_0-2,\cdots,j_2^* + 1\}$, until either $x'_{2,j_2^*}=0$ or ${J_2}=\{j_2^* + 1,\cdots,N\}$. In either case, for any ${j_1} \in {J_1}$ and ${j_2} \in {J_2}$, there will be $j_1<j_2$.

Furthermore, at least $N-1$ UEs will belong to $J_1$ or $J_2$ (Lemma 1), if $J_1\bigcup{J_2}=\{1,\cdots,N\}$, $j_{1,2}=\max\limits_{j\in{J_1}}\{j\}$. If UE $j^*$ is powered by both BS 1 and BS 2 simultaneously, $j_{1,2}={j^*}$.

\subsection{$M=3$, $\pmb{J^{CC}}=\left[{j_{1,2}},{j_{1,3}},j_{2,3}\right]$}
\begin{figure}[!htb]
\centering
\includegraphics[width=2in]{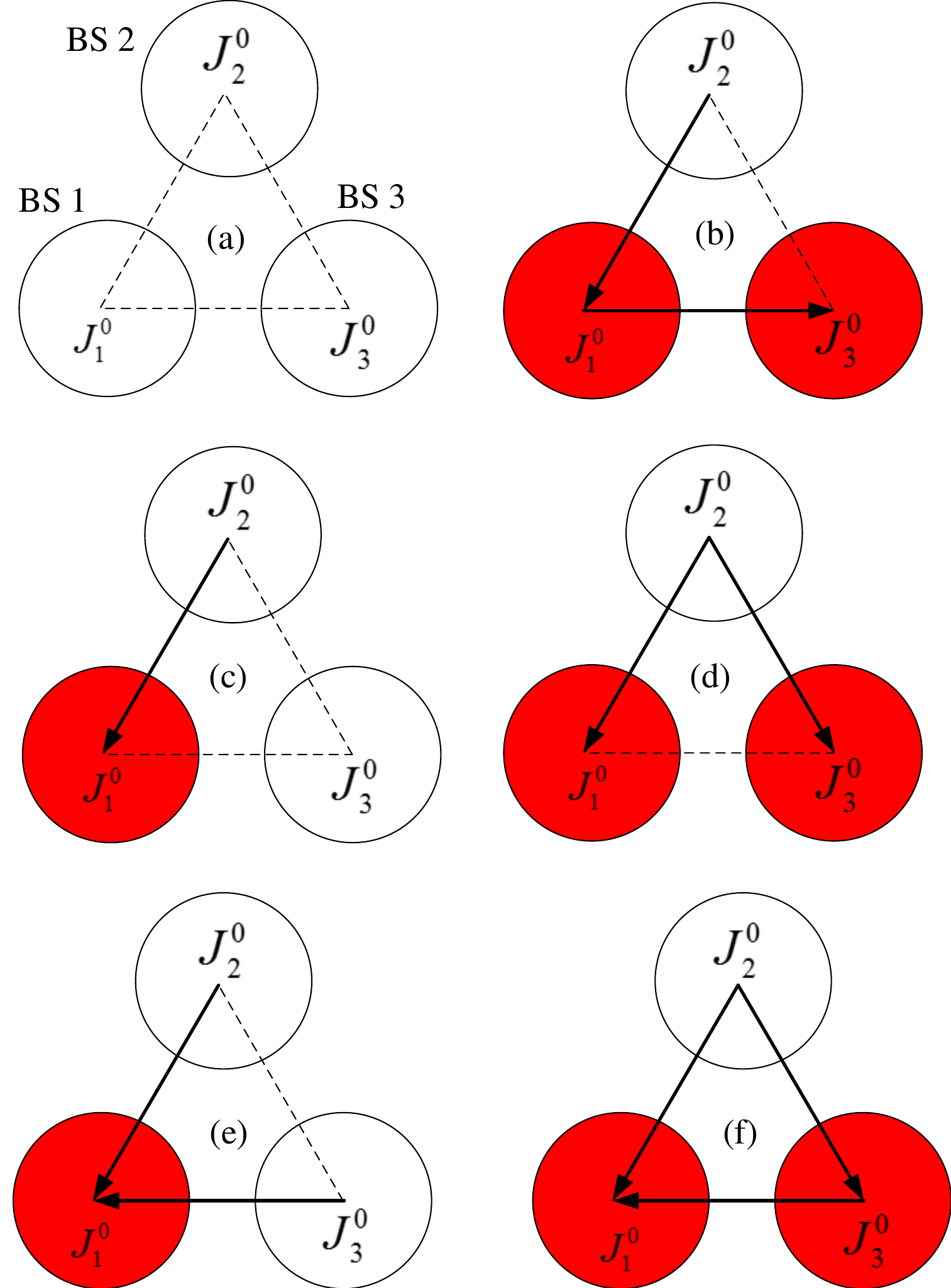}
\caption{Spectrum and power coordination in the 3-BS cooperative wireless system (red BS means power consumption is 1).}
\label{association3BS}
\end{figure}

The UE-BS association scheme in the optimal solution must fall in one of the categories in Fig. \ref{association3BS}, where a solid arrow represents power coordination (providing power for other BS's UE), and a dashed line means spectrum coordination only (increasing the transmission power of its own UE to spare more spectrum for other BSs).

1) Fig. \ref{association3BS} (a):  $J_i=J_i^0$, and for initial clusters $J_{i_1}^0$ and $J_{i_2}^0$, let $j_{i_1,i_2}=j_{i_1,i_2}^0$.

2) Fig. \ref{association3BS} (e)-(f): Suppose BS $1$ is receiving power coordination from both BS $2$ and BS $3$, then BS $1$ cannot provide power for UE belonging to $J_2^0$ or $J_3^0$.
\[J_1\subseteq{J^0_1}\]

\begin{figure}[!htb]
\centering
\includegraphics[width=2in]{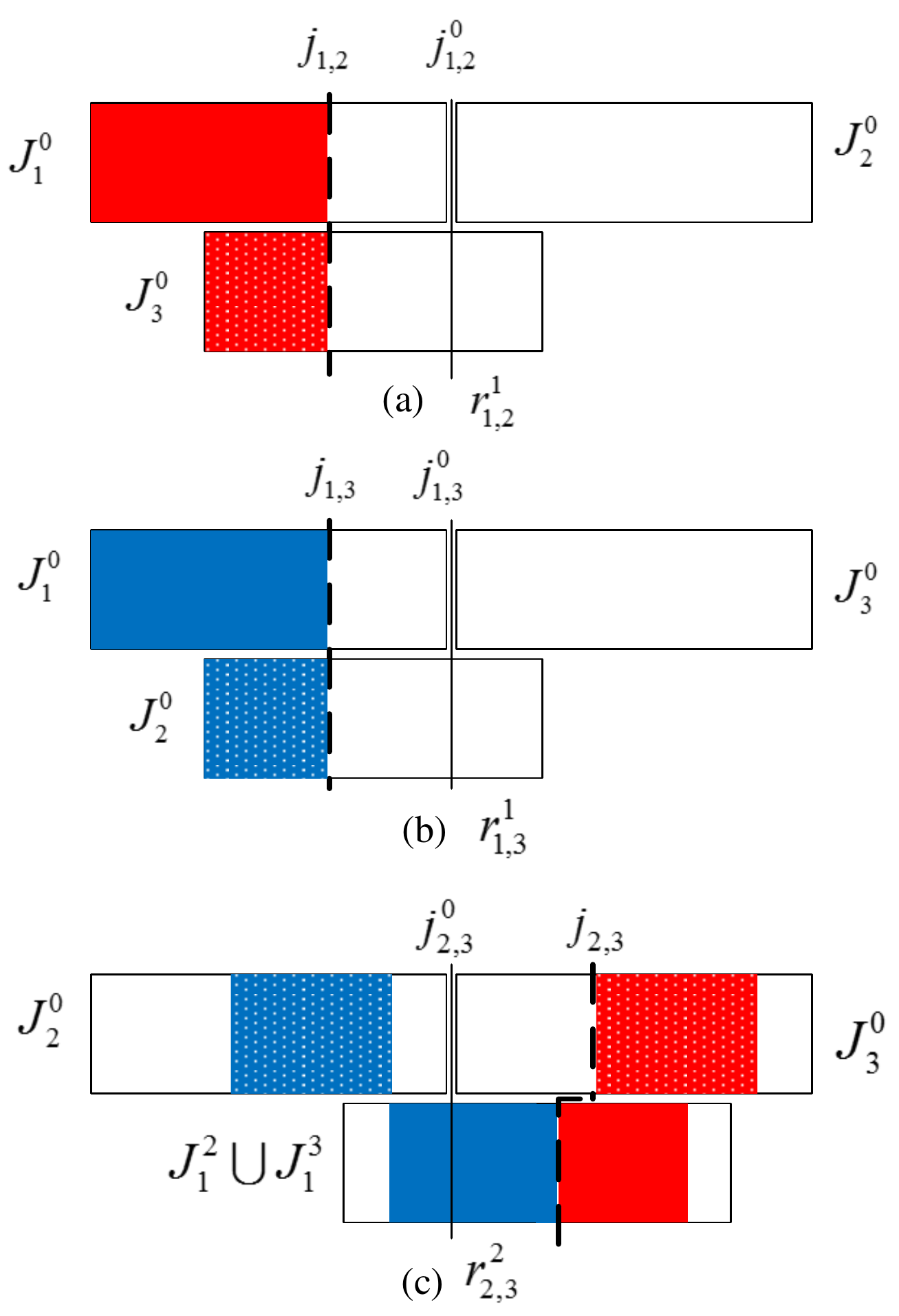}
\caption{UE-BS association in the 3-BS cooperative wireless system (Fig. \ref{association3BS} (e), (f)).}
\label{association3BS2}
\end{figure}

Suppose UEs are sorted in descending order of ${\gamma^j_{1,2}}$, ${\gamma^j_{1,3}}$, and ${\gamma^j_{2,3}}$ respectively, as shown in Fig. \ref{association3BS2}. In order to allow BS $1$ receive power coordination from both BS $2$ and BS $3$, we have
\[{j_{1,2}}\le{j^0_{1,2}}\text{, }{j_{1,3}}\le{j^0_{1,3}}\]

According to (\ref{con}), UEs outside the red or blue region in Fig. \ref{association3BS2} (a)-(b) do not belong to $J_1$. Similarly, UEs in the red region do not belong to $J_2$ and UEs in the blue region do not belong to $J_3$. Then, we can conclude $J_1$ is the intersection of of red and blue regions in $J^0_1$.

Since ${J_1}\bigcup{J^2_1}\bigcup{J^3_1}\bigcup{J^0_2}\bigcup{J^0_3}=\{1,\cdots,N\}$, Fig. \ref{association3BS2} (c) is similar to Fig. \ref{association}, with the difference that in the cluster ${J^2_1}\bigcup{J^3_1}$, at most two UEs will be powered by BS 1 (one in blue region and one in red region). The corresponding $j_{2,3}$ can be designed according to (\ref{con}).

\emph{Note 2}: Suppose UE $j^*\in {J^0_1}$ is powered by three BSs simultaneously, the red and blue regions will overlap with UE $j^*$ in Fig. \ref{association3BS2} (c). Since UEs other than $j^*$ will be powered by individual BSs (Lemma 1), we will have
\[j_{1,2}=j_{1,3}=j_{2,3}={j^*}\]

3) Fig. \ref{association3BS} (b)-(d): There exists UE $j$, $j\in{J^0_1}$ receiving power from BS $2$, while no UE in ${J^0_1}$ will receive power from BS $3$. So,
\[{j_{1,2}}\le{j^0_{1,2}}\text{, }{j_{1,3}}\ge{j^0_{1,3}}\]

\begin{figure}[!htb]
\centering
\includegraphics[width=2in]{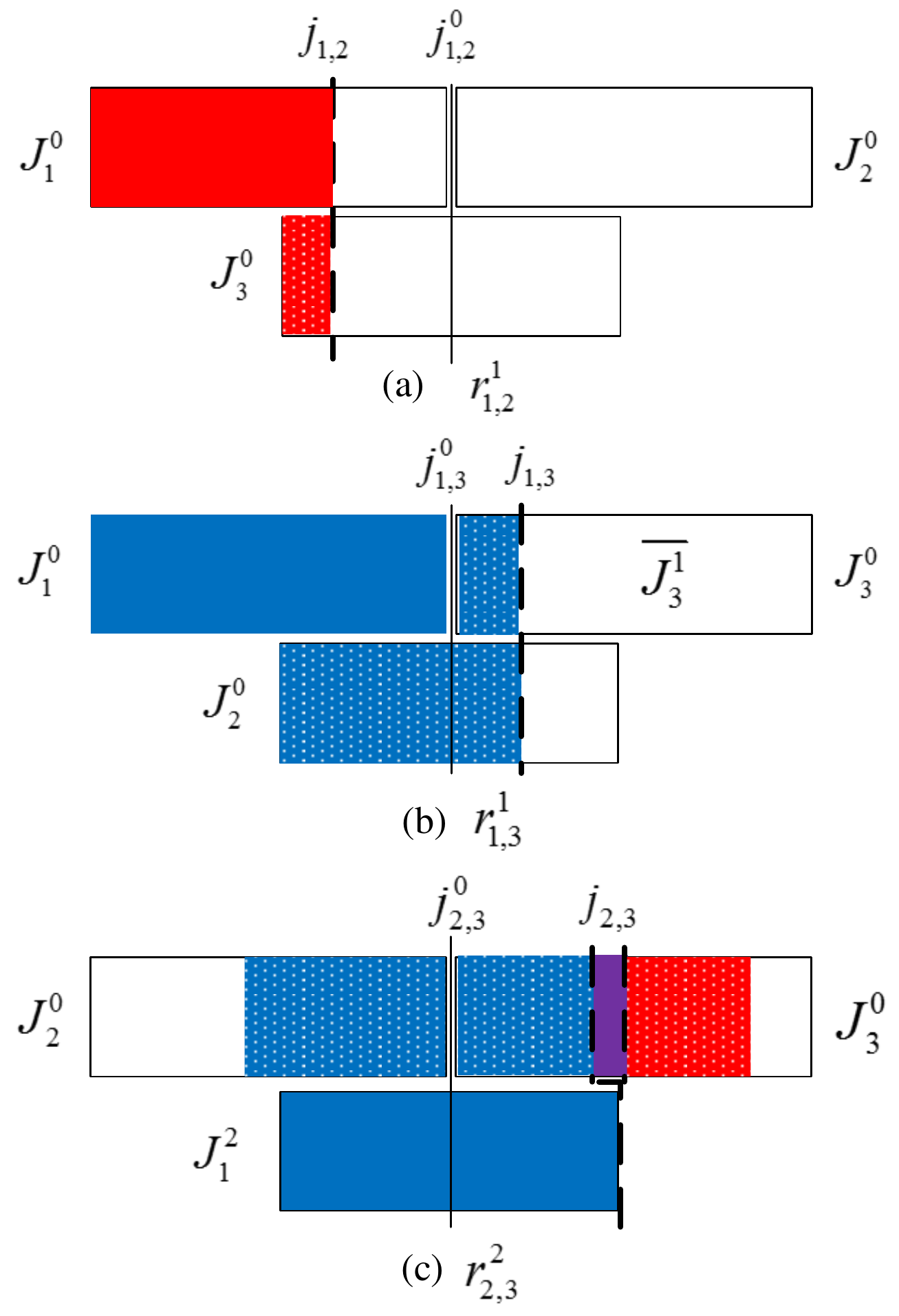}
\caption{UE-BS association in the 3-BS cooperative wireless system (Fig. \ref{association3BS} (b)-(d)).}
\label{association3BS3}
\end{figure}

According to (\ref{con}), in Fig. \ref{association3BS3} (a)-(b), the intersection of red and blue regions of $J^0_1$ belongs to $J_1$, and so does the intersection of the red and blue region of $J^0_3$, which is indicated as the purple region in Fig. \ref{association3BS3} (c).

Meanwhile, $J_2$ will not include UEs which are located in the red region of Fig. \ref{association3BS3} (c), and $J_3$ will not include UEs which are located in the blue region. So, ${j_{2,3}}\ge{j^0_{2,3}}$ will be a value that is able to separate these two regions. 


\section{Proof of Lemma 3}\label{c}
If $S_1=S_2$, then they must be the optimal solution because relaxing the constraints by limiting the transmit power of only one BS brings no benefits to the total power consumption.

Then, we only prove part 2) with $S_{1}>({1,1})$, as part 3) can be similarly proved. Suppose in the CC vector $\pmb{J^{CC}}$ which corresponds to the optimal solution $\{x^*_{i,j}\}$, we have $j_{1,2}< j'_1$, then the optimal cluster $J_1^* \subseteq \{ 1,\cdots,j'{_1} - 2\}\subset{J_1^0}$. 

\begin{figure}[!t]
\centering
\vspace{-1.5em}
\includegraphics[width=3.2in]{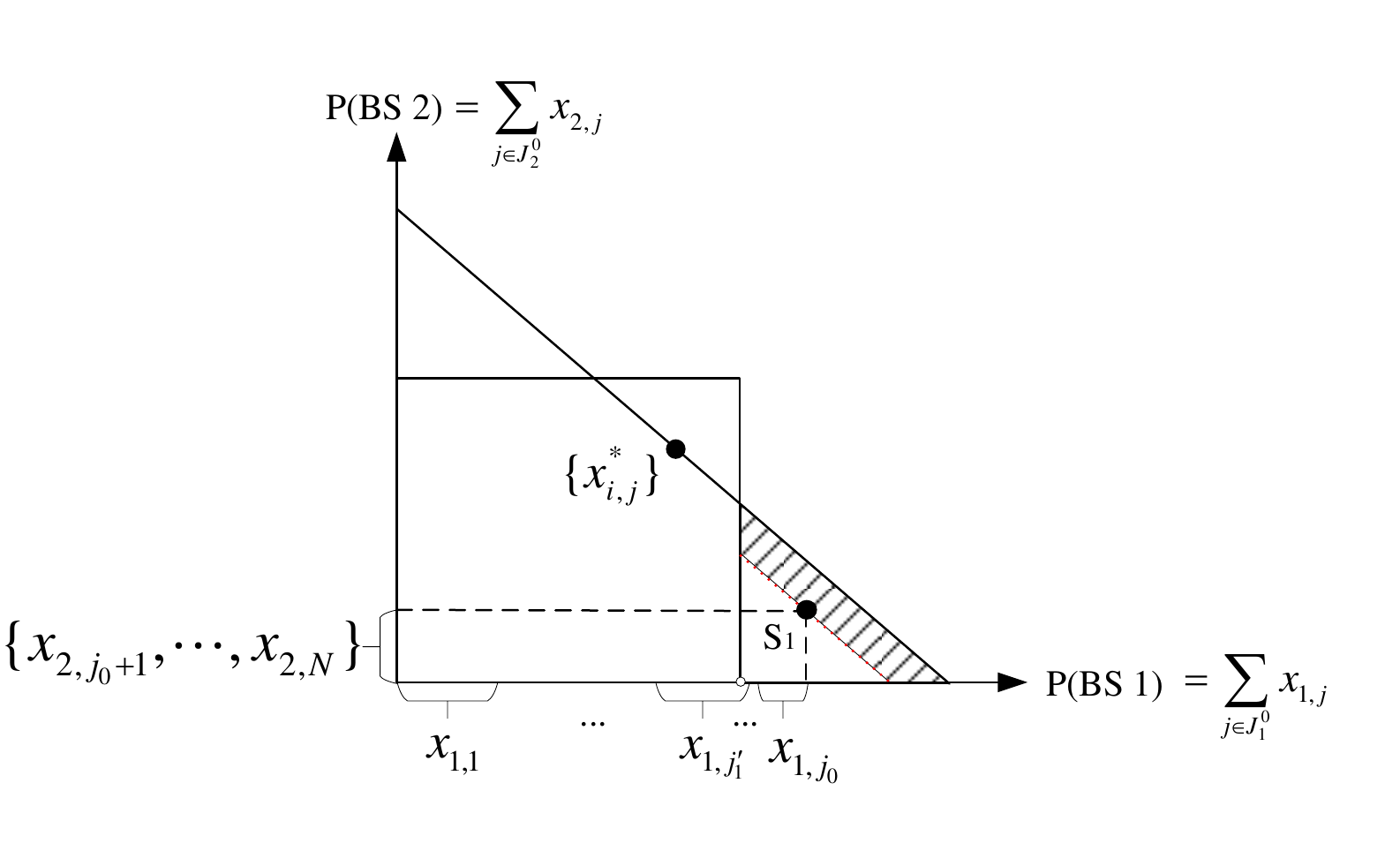}
\vspace{-1em}
\caption{Solution Mapping, straight slop=1.}
\label{map1}
\end{figure}

1) For the optimal solution $\{x^*_{i,j}\}$, we can find the corresponding mapping $\{{x'_{i,j}}\}$ by relaxing the constraint of BS 1 as follows:
\begin{equation}\label{add}
\begin{array}{*{20}{l}}
{x'_{1,j}}=x^*_{1,j}+{\gamma_{2,1}^j}{x^*_{2,j}},&x'_{2,j}=0,&j\in{J_1^0}\backslash{J^*_1}\\
x'_{1,j}=x^*_{1,j},&x'_{2,j}=x^*_{2,j},&j\in {J^*_1}\bigcup{J_2^0}
\end{array}
\end{equation}
Since $\{{x_{i,j}}\}$ in $S_1$ is the optimal relaxed solution, $\{{x'_{i,j}}\}$ must fall into the shadowed region of Fig. \ref{map1},
\begin{equation}\label{third}
\sum\limits_{j \in {J_1^0}} {x'_{1,j}}+\sum\limits_{j \in {J_2^0}} {x'_{2,j}}>\sum\limits_{j \in {J_1^0}} x_{1,j}+\sum\limits_{j \in {J_2^0}} {x_{2,j}}
\end{equation}
Based on (\ref{third}), we have the following result
\begin{equation}\label{forth}
\sum\limits_{j = 1}^{{ j{'_1} -1}}\left({{x'_{1,j}}}-{{x_{1,j}}}\right)>\sum\limits_{j = j{'_1}}^{j{_0}} \left(x_{1,j}- x'_{1,j}\right)+\sum\limits_{j \in {J_2^0}} \left({x_{2,j}}- {x'_{2,j}}\right)
\end{equation}

2) For $\{x_{i,j}\}$, the relaxed solution in $S_1$, we can find the corresponding mapping $\{{x^{*'}_{i,j}}\}$ by considering the power constraint of BS 1. The power allocation with power constraints for each BS becomes
\begin{equation}\label{fifth}
\begin{array}{*{20}{l}}
x^{*'}_{1,j}=x_{1,j},\:x^{*'}_{2,j}=0,\:j\in \{1,\cdots,j'_1-1\}\\
{x^{*'}_{1,j}}={\beta} {x_{1,j}},\:x^{*'}_{2,j}={\gamma_{1,2}^j}(1-\beta){x_{1,j}},\:j=j'_1\\
x^{*'}_{1,j}=0,\:x^{*'}_{2,j}={\gamma_{1,2}^j}{x_{1,j}},\:j\in \{j'_1+1,\cdots,j_0\}\\
x^{*'}_{1,j}=0,\:x^{*'}_{2,j}=x_{2,j},\:j\in J_2^0
\end{array}
\end{equation}
where $\beta = (1-\sum\nolimits_{j = 1}^{{j{'_1}-1}} {{x_{1,j}}})/{x_{1,j{'_1}}}$.
From (\ref{add}) and (\ref{fifth}), we have
\begin{equation}\label{add1}
\begin{array}{l}
\sum\limits_{i = 1}^2\sum\limits_{j = 1}^{{ j{'_1} -1}}\left({x^*_{i,j}}-x'^*_{i,j}\right)\ge\\
\left({\sum\limits_{j = 1}^{{ j{'_1} -1}}{{x^*_{1,j}}}}+({\sum\limits_{j = 1}^{{ j{'_1} -1}}{{x'_{1,j}}}}-{\sum\limits_{j = 1}^{{ j{'_1} -1}}{{x^*_{1,j}}}}){\gamma_{1,2}^{j'-1}}\right)-\sum\limits_{j = 1}^{{ j{'_1} -1}}{{x_{1,j}}}\\
\sum\limits_{i = 1}^2\sum\limits_{j = j{'_1}}^{{ N}}\left(x'^*_{i,j}-{x^*_{i,j}}\right)=\left((1-{\gamma_{1,2}^{j'_1}}){\beta}{x_{1,j'_1}}+\right.\\
\left.\sum\limits_{j = j{'_1}}^{j{_0}} {\gamma_{1,2}^j}x_{1,j}+\sum\limits_{j \in {J_2^0}} {x_{2,j}}\right)-\left(\sum\limits_{j = j{'_1}}^{j{_0}}{\gamma_{1,2}^j} x'_{1,j}+\sum\limits_{j \in {J_2^0}} {x'_{2,j}}\right)
\end{array}
\end{equation}

Substituting $\sum\nolimits_{j = 1}^{ j{'_1} -1}{{x^*_{1,j}}}=1=\sum\nolimits_{j = 1}^{ j{'_1} -1}{{x_{1,j}}}+{\beta} {x_{1,j'_1}}$ into (\ref{add1}), we can see
\begin{equation}\label{sixth}
\begin{array}{l}
\sum\limits_{i = 1}^2\sum\limits_{j = 1}^{{ j{'_1} -1}}\left({x^*_{i,j}}-x'^*_{i,j}\right)\ge\beta {x_{1,j'_1}}+\\
\left(\sum\limits_{j = 1}^{{ j{'_1} -1}}\left({{x'_{1,j}}}-{{x_{1,j}}}\right)-\beta {x_{1,j{'_1}}}\right){\gamma_{1, 2}^{j{'_1}-1}}\\

\sum\limits_{i = 1}^2\sum\limits_{j = j{'_1}}^{{ N}}\left(x'^*_{i,j}-{x^*_{i,j}}\right) \le\beta {x_{1,j'_1}}+\\ 
\left(\sum\limits_{j = j{'_1}}^{j{_0}} \left(x_{1,j}- x'_{1,j}\right)-\beta {x_{1,j{'_1}}}\right){\gamma_{1, 2}^{j{'_1}}}+\sum\limits_{j \in {J_2^0}} \left({x_{2,j}}- {x'_{2,j}}\right)\\
\end{array}
\end{equation}

According to the sorting rule of UE, ${\gamma_{1, 2}^{j{'_1}-1}}\ge{\gamma_{1, 2}^{j{'_1}}}$. From (\ref{forth}) and (\ref{sixth}), we have
\[\sum\limits_{i = 1}^2\sum\limits_{j = 1}^{{ j{'_1} -1}}\left({x^*_{i,j}}-x'^*_{1,j}\right)\ge\sum\limits_{i = 1}^2\sum\limits_{j = j{'_1}}^{{ N}}\left(x'^*_{1,j}-{x^*_{i,j}}\right)\]
Consequently, the following result can be obtained which contradicts with the assumption that $\{x^*_{i,j}\}$ is the optimal solution.
\[\sum\limits_{i = 1}^2\sum\limits_{j = 1}^N {x^*_{i,j}}\ge\sum\limits_{i = 1}^2\sum\limits_{j = 1}^N {x'^*_{i,j}}\]

%
%



%

\bibliographystyle{unsrt}
\bibliography{mybib}
\end{document}